\documentclass[trackchanges,twocolumn,twocolappendix]{aastex701}

\usepackage{amsmath}
\usepackage{bm}
\usepackage{float}

\begin{document}

\title{First application of weak lensing peak steepness statistics to HSC Y1 data: \\
effectively probing halo density profiles}

\author[orcid=0000-0001-7140-1950]{Ziwei Li}
\affiliation{South-Western Institute for Astronomy Research, Yunnan University, Kunming 650500, China}
\email{lzw@ynu.edu.cn}  

\author[orcid=0000-0003-0394-1298]{Xiangkun Liu}
\email[show]{liuxk@ynu.edu.cn}
\affiliation{South-Western Institute for Astronomy Research, Yunnan University, Kunming 650500, China}
\affiliation{Yunnan Key Laboratory of Survey Science, Yunnan University, Kunming 650500, China}

\author[orcid=0000-0002-0876-9143]{Tianyu Zhang}
\affiliation{South-Western Institute for Astronomy Research, Yunnan University, Kunming 650500, China}
\affiliation{School of Physics and Astronomy, Beijing Normal University, Beijing 100875, China}
\email{zhangty@bnu.edu.cn}

\author{An Zhao}
\affiliation{South-Western Institute for Astronomy Research, Yunnan University, Kunming 650500, China}
\email{zhaoan@mail.ynu.edu.cn}

\author{Chuzhong Pan}
\affiliation{Department of Astronomy, School of Physics, Peking University, Beijing 100871, China}
\email{panchuzhong@126.com}

\author{Shuo Yuan}
\affiliation{National Astronomical Observatories, Chinese Academy of Sciences, Beijing 100101, China}
\email{yuanshuoastro@gmail.com}

\author[orcid=0000-0003-2153-7758]{Qiao Wang}
\affiliation{National Astronomical Observatories, Chinese Academy of Sciences, Beijing 100101, China}
\affiliation{School of Astronomy and Space Science, University of Chinese Academy of Sciences, Beijing 100048, China}
\email{qwang@nao.cas.cn}

\author[orcid=0000-0002-8397-012X]{Zuhui Fan}
\email[show]{zuhuifan@ynu.edu.cn}
\affiliation{South-Western Institute for Astronomy Research, Yunnan University, Kunming 650500, China}
\affiliation{Yunnan Key Laboratory of Survey Science, Yunnan University, Kunming 650500, China}
\affiliation{Southwest United Graduate School, Kunming 650092, China}


\begin{abstract}

As a new probe, the weak lensing  (WL) peak steepness statistics is sensitive to the density profile of halos that encodes important
information of baryonic feedback and dark matter properties, leading to a promising means to statistically constrain these effects using WL data.
In this article, we present its first application to HSC Y1 data to demonstrate the great potential of this new statistics. 
Within the phenomenological framework of HMcode2016 that attributes the baryonic feedback solely to the reduction of the halo concentration parameter and focusing on high peaks originated dominantly from massive clusters, 
our analyses by combining WL peak height and steepness statistics resulted in $S_8=0.76^{+0.08}_{-0.07}$ with the maximum-a-posteriori (MAP) of $0.79$ and low concentrations. 
Taking the form of the concentration-mass relation as $c(M,z)=A(1+z_{\rm f})/(1+z)$ with $z_{\rm f}$ being the formation redshift of halos with mass $M$ at redshift $z$, we obtain  
$A=1.93^{+1.33}_{-1.16}$ (MAP=$1.70$) in comparison with $A=3.34^{+1.52}_{-1.74}$ (MAP=3.31) from dark matter only simulated mocks.  
The result tends to support phenomenologically strong baryonic feedback effects at cluster scales.

\end{abstract}



\section{Introduction}

With the fast developments of large photometric surveys \citep[e.g.,][]{2012MNRAS.427..146H, 2016MNRAS.460.1270D, 2015MNRAS.454.3500K, 2018PASJ...70S...4A}, weak lensing (WL) has become 
a major cosmological probe in mapping directly the large-scale mass distribution in the Universe \citep[e.g.,][]{2001PhR...340..291B,2014RAA....14.1061F,2015RPPh...78h6901K,2015IJMPD..2430011F,2018ARA&A..56..393M},
resulted in significant cosmological constraints that are comparable and importantly complement to the information derived from other probes \citep[e.g.,][]{2021A&A...646A.140H,2022PhRvD.105b3520A,2023PhRvD.108l3517M,2023PhRvD.108l3521S,2025arXiv250319441W,2025arXiv250903582A}. With the Stage IV surveys being in place \citep[]{2019ApJ...873..111I, 2011arXiv1110.3193L,2025A&A...697A...1E, 2019arXiv190205569A, 2019ApJ...883..203G,2023A&A...669A.128L}, we are now entering 
a new era of WL cosmological studies with greatly improved data in quantity and quality.    

Along with the observational advance, the methodologies to extract physical information from WL data have also been developed. Because of the non-Gaussianity of the cosmic matter distribution, 
different statistics beyond two-point correlations (2pt)
have been investigated and applied to existing surveys \citep[e.g.,][]{2013MNRAS.433.3373V,2014MNRAS.441.2725F,2022PhRvD.105j3537S,2023PhRvD.108l3526T,2025arXiv250814018G,2025arXiv250814019S,
2015PhRvD..92f3517L,2015MNRAS.450.2888L,2016MNRAS.463.3653K,2018MNRAS.474.1116S,2018MNRAS.474..712M,2022MNRAS.511.2075Z,2023MNRAS.519..594L,2024MNRAS.534.3305H,2024MNRAS.528.4513M,2025JCAP...01..006C,2025PhRvD.111h3510N,2025MNRAS.537.3553A,2026MNRAS.546ag033C}. 
Utilizing different statistics to enhance cosmological gains has become one of the important efforts for Stage IV surveys \citep[e.g.,][]{2023A&A...675A.120E,2025arXiv251004953E}. 
Apart from the summary statistics, machine-learning based methods dealing with field-level data have drawn increasing attention from framework studies to observational applications \citep[e.g.,][]{2018PhRvD..97j3515G,2019NatAs...3...93R,
2019MNRAS.490.1843R,2019PhRvD.100f3514F, 2020PhRvD.102l3506M,2021MNRAS.504.1825S,2022PhRvD.105h3518F, 2023MNRAS.521.2050L, 2024PhRvD.110d3535Z,2024PNAS..12109624D,2025A&A...697A.162L,2025MNRAS.536.1303J}.  

Among different analyses, WL peak statistics has long been recognized as an efficient probe to extract non-Gaussian information from WL data \citep[e.g.,][]{2004MNRAS.350..893H, 2009ApJ...698L..33M, 2010MNRAS.402.1049D,
2010PhRvD..81d3519K, 2010ApJ...719.1408F, 2011ApJ...728L..13M, 2011PhRvD..84d3529Y, 2013MNRAS.432.1338M,2015A&A...576A..24L,2015A&A...581A.101M}.
Up to now, the majority of the peak analyses concentrate on peak height statistics \citep[e.g.,][]{2015MNRAS.450.2888L,2016MNRAS.463.3653K,2018MNRAS.474.1116S,2018MNRAS.474..712M,2022MNRAS.511.2075Z,
2023MNRAS.519..594L,2024MNRAS.534.3305H}. In the machine learning studies of \citep[]{2019NatAs...3...93R}, by analyzing the extracted features, it is revealed that the statistics in terms of the peak profile, refer to as peak steepness, 
can provide more cosmological information than that from the peak height statistics. Inspired by this, in \cite{2023MNRAS.520.6382L}, we perform systematic studies to compare the two peak statistics with N-body simulations, 
and extend our theoretical model for high peak height statistics \citep{2010ApJ...719.1408F, 2018ApJ...857..112Y} to the corresponding steepness statistics.  
We find that in the case of low shape noise (high source number density) expected from Stage IV WL surveys, the peak steepness statistics indeed can tighten the cosmological constraints comparing  
with peak height statistics. Importantly, with the sensitivity to the halo profile, the steepness statistics can effectively probe the inner mass distribution of halos. 
The latter contains rich information of astrophysics and the nature of dark matter, further enhancing the advantage of including peak steepness statistics in WL analyses.       

Here we present the peak steepness analyses using real observational data for the first time to demonstrate its feasibility and great potential in WL studies. 

In Sec.~\ref{sec:data}, we summarize the observational data used in the analyses. The simulation mocks are described in Sec.~\ref{sec:mocks}. Sec.~\ref{sec:Peak model} presents
our theoretical model for high peaks.
The mock validation results are shown in Sec.~\ref{sec:Mock validation}. Potential systematic effects are discussed in Sec.~\ref{sec:systematics}. Sec.~\ref{sec:Observational results} contains the results from
the observational data. Summary and discussion are shown in Sec.~\ref{sec:Summary and Discussion}.

\section{Observational data}\label{sec:data}

In our observational analysis, we use the first-year (Y1) shear catalog from the Hyper Suprime-Cam Subaru Strategic Program (HSC-SSP) (S16A) \citep[]{2018PASJ...70S..25M}, which was also employed in our previous tomographic peak height studies 
in \cite{2023MNRAS.519..594L}. Specifically, we select the galaxy shear sample following the criteria listed in Table 4 of \cite{2018PASJ...70S..25M} and limit their redshift range to be $0.2\le z_{\rm p}\le 1.5$ as suggested by
the HSC photometric redshift measurement studies of \cite{2018PASJ...70S...9T} ,where $z_{\rm p}$ is the best-fitted value of the photometric redshift (photo-z) of galaxies.

From S16A, we choose $52$ regular square fields for our peak analyses, which consist of 40 fields with an area of $1.5\times1.5\ \deg^2$ each and 12 fields with $1.0\times1.0\ \deg^2$ each.
This yields a total area of $102\ \deg^2$. For each selected area, we first calculate the shear field on a grid of $1024\times 1024$ from the galaxy shear sample \citep[]{2018PASJ...70S..26O} by applying 
a smoothing kernel with $W(|\bm{\theta}|)=\exp \left(-|\bm{\theta}|^{2}/\theta_{\rm G}^{2}\right)/\pi \theta_{\rm G}^{2}$ taking $\theta_{\rm G}=1.5 \hbox{ arcmin}$. We then 
perform the convergence reconstruction (denoted as $K_{\rm N}$) using the non-linear Kaiser-Squires inversion \citep[]{1993ApJ...404..441K,1997A&A...318..687S,2015MNRAS.450.2888L}. 
By randomly rotating galaxies to eliminate shear signals, we also generate noise maps following the same procedures. In addition, we create filling factor maps from the spatial distribution of galaxies to identify masked
regions in our peak analyses. 

We note that the pixel size for a field of $1.5\times1.5\ \deg^2$ and of $1.0\times1.0\ \deg^2$ is somewhat different
with the value of $0.088\hbox{ arcmin}$ and $0.059\hbox{ arcmin}$, respectively. We take the specific size value into
the calculations of the peak steepness. Since both are much smaller than the smoothing scale, the different pixel size
for the two sets of maps has no significant impacts on our analyses \citep{2023MNRAS.520.6382L}.

For the KS reconstruction, previously we investigated extensively the mask and boundary effects on the reconstructed convergence fields and subsequently on the peak statistics
\citep[]{2014ApJ...784...31L}.
These studies ensure us the use of the KS reconstruction in our analyses here.
To control the systematic impacts from masks and boundaries on the high peak statistics, we remove the outer most $5\theta_{\rm G}$ regions along each side of a map
and the regions with the filling factor $f<0.6$ for peak counting.
The resulting total effective area is $\sim 57.4\deg^2$. Further details can be found in \cite{2023MNRAS.519..594L}.

Different convergence reconstruction methods have been proposed and applied in different analyses
aiming for, e.g., mitigating the mask and boundary effects with inpainting techniques, and reducing the shape noise effects with nonlinear regularizations
\citep[e.g.,][]{2018MNRAS.479.2871J,2014MNRAS.440.1281L,2020A&A...638A.141P, 2021A&A...649A..99S,2022MNRAS.512...73F,2024PhRvD.109l3530S}.
Nonlinear regularizations lead to highly non-Gaussian shape noise fields, making our theoretical peak model inapplicable \citep[e.g.,][]{2011RAA....11..507J}.
On the other hand, the inpainting techniques can significantly reduce the mask and boundary effects increasing the effective survey areas. However, before utilizing such reconstructions,
we still need to investigate carefully the possible systematics, which is one of our future tasks.

\begin{figure*}
\centering
\includegraphics[width=2.0\columnwidth]{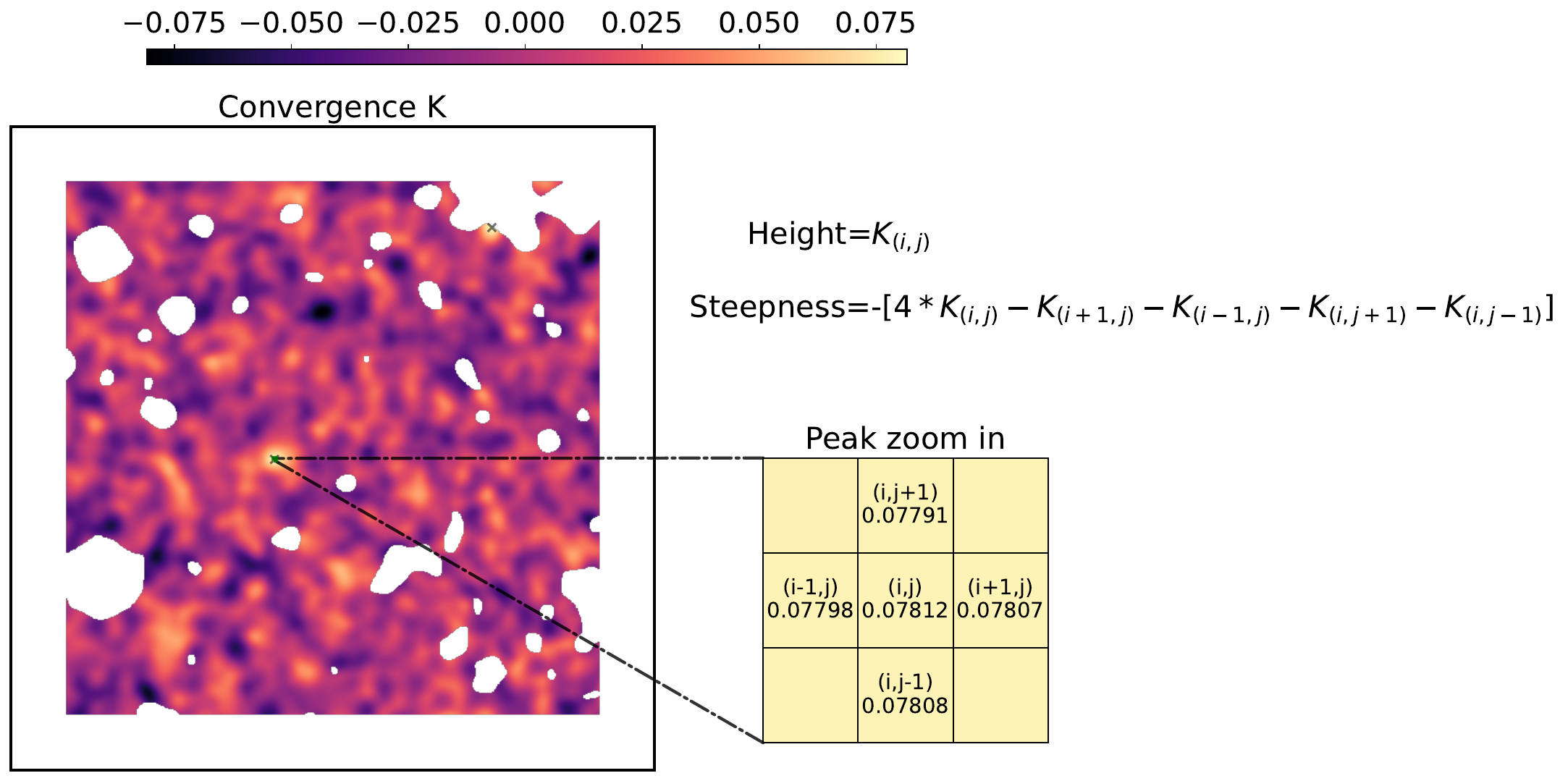}
\caption{\label{fig:demo} An example of a convergence map reconstructed from the HSC S16A data with the excluded masks and the boundaries shown in white. Peaks with $\nu\ge 4$ 
are labeled with cross symbols. The zoom-in part shows the peak identification and its steepness calculation. 
}
\end{figure*}

Figure \ref{fig:demo} shows an example of a reconstructed convergence map with the peaks of $\nu\ge 4$ labeled by cross symbols 
and the excluded regions shown in white, where 
$\nu=K_{\rm N}/\sigma_{\rm N,0}$ with $\sigma_{\rm N,0}$ being the root mean square (rms) of the randomly rotated noise maps and $\sigma_{\rm N,0}\approx 0.018$ for the S16A data. A peak is defined as the pixel with its $K_{\rm N}$ value higher than its neighboring $8$ pixels. To calculate the steepness of a peak, we follow \cite{2023MNRAS.520.6382L} and use the discrete operator $[S]=4\times[K_{\rm N} \hbox{ at the peak position}]-[\hbox{Sum of } K_{\rm N} \hbox{ of the nearest 4 pixels}]$, which corresponds to $-(K_{\rm N}^{11}+K_{\rm N}^{22})$ with $K_{\rm N}^{ij}$ being the second derivative of $K_{\rm N}$ with respect to $i,j$ coordinates. We demonstrate the steepness calculation in the zoom-in part of Figure \ref{fig:demo}. The signal-to-noise ratio of the steepness of a peak is defined as $x=S/\sigma_{\rm N,2}$, where $\sigma_{\rm N,2}$ is the rms of the second moment of the noise maps and $\sigma_{\rm N,2}\approx 0.022\hbox{ arcmin}^{-2}$ here.

\section{Simulation mocks}\label{sec:mocks} We adopt the same sets of simulation mocks used in \cite{2023MNRAS.519..594L} to validate our analyzing method and calculate the covariance matrix. Briefly, 
the mocks are based on $24$ sets of N-body simulations assuming flat $\Lambda$CDM with the present matter density, dark energy density in the form of cosmological constant, baryon density, Hubble constant
in unit of $100\hbox{ km/s/Mpc}$ and the power index of primordial matter density perturbations being $(\Omega_{\rm m},\Omega_\Lambda,\Omega_b,h,\sigma_8,n_s$) = ($0.28,0.72, 0.046,0.7,0.82,0.96$). Each set 
includes $8$ and $4$ simulations with the respective box size of $320h^{-1}\hbox{Mpc}$ and $600h^{-1}\hbox{Mpc}$ and all have the particle number of $640^3$. The WL ray tracing calculations are done
for each set using the $8$ smaller ones to fill the light cone to $z=1$ and the $4$ larger ones for $z=1$ to $z=3$ with a total of $59$ lens planes, generating $4$ sets of independent 
lensing maps each with an area of $3.5\times 3.5 \deg^2$ sampled on $1024\times 1024$ pixels. In total, we have $24\times 4=96$ sets of lensing maps.   

To generate S16A mocks, for an observational field, we keep the position and the redshift $z_{\rm p}$ of the observed galaxies and remove their shear signals by random rotation. 
We then use the simulated maps to generate mock WL shear signals on the galaxies by interpolating in the angular and redshift dimensions. 
Following \cite{2018PASJ...70S..26O}, the intrinsic ellipticities of galaxies and the shear measurement errors are accounted for
in building the mock `observed' ellipticity for each galaxy \citep{2023MNRAS.519..594L}. 
For each of the $52$ selected fields, we generate $20$ sets of mocks each with different mock WL signals based on a different set of simulated WL maps chosen from the $96$ sets.
We then construct a S16A mock by randomly selecting one from the $20$ sets for each field and totally generate $1000$ bootstrap mocks. 
The same convergence reconstruction and the analyzing procedures as those dealing with observational data are applied to 
these mocks. Finally, the average peak counts in terms of height and steepness and the corresponding covariance are calculated based on the $1000$ mocks. 

For the mock convergence maps, the pixel sizes are the same as the corresponding observational maps. However,
we need to mention that the mock WL signals are assigned based on the ray-tracing maps with the pixel size of
$3.5\times 60/1023\approx 0.2\hbox{ arcmin}$, larger than the pixel size of the reconstructed mock convergence maps. With the much larger smoothing scale $\theta_G=1.5\hbox{ arcmin}$ applied
in our analyses, the effects from the different pixel sizes here are negligible.

\section{Peak model}\label{sec:Peak model}
Simulation and observational studies have shown that for WL high peaks, they are mainly associated with massive halos of cluster scale \citep[e.g.,][]{2018ApJ...857..112Y, 2025MNRAS.538..755B,
2024OJAp....7E..90C,2025OJAp....8E...2C}.
For lower peaks, they are largely from the cumulative contributions of the line-of-sight large-scale structures without dominant halos. In our analyses, we utilize our halo-based peak model for cosmological inferences
that is valid for high peaks originated dominantly from massive halos. We therefore consider peaks with $\nu\ge 4$ corresponding to $K_{\rm N}\gtrsim 0.072$ here.

Our theoretical model for high peak height statistics is presented in \cite{2010ApJ...719.1408F, 2018ApJ...857..112Y}, and is later extended to the peak steepness statistics
showing explicitly its sensitivity to the halo density profile in \cite{2023MNRAS.520.6382L}. In this model, $K_{\rm N}$ is written as 
\begin{equation}\label{eq:K_N model}
  K_{\rm N}=K_{\rm H}+K_{\rm{LSS}}+N,
\end{equation}
where $K_{\rm H}$ is the contribution from massive halos with mass $M$ larger than a lower limit $M_*$, $K_{\rm{LSS}}$ represents the large-scale projection effect other than the massive halos included in $K_{\rm H}$, and $N$ denotes the shape noise field.
Under the assumption that both $K_{\rm{LSS}}$ and $N$ can be approximated as Gaussian random fields, the abundance of peaks in a given $K_{\rm H}$ halo region can be calculated by the Gaussian random field theory 
\citep{1986ApJ...304...15B,1987MNRAS.226..655B} modulated by $K_{\rm H}$. 
The cosmological information related to $K_{\rm H}$ is the halo mass function and the halo density profile. 
For $K_{\rm{LSS}}$, the quantities of its zero, first and second moments $\sigma_{\rm{LSS},0}$, $\sigma_{\rm{LSS},1}$ and $\sigma_{\rm{LSS},2}$ enter the calculations, and they are all determined by the 
power spectrum of $K_{\rm{LSS}}$. For the shape noise field $N$, without considering IA effects, their moments $\sigma_{\rm N,0}$, $\sigma_{\rm N,1}$ and $\sigma_{\rm N,2}$ are cosmology-free.  
In Appendix \ref{app:model}, we present some theoretical details of the model.

Within this theoretical framework, different astrophysical effects can be incorporated as long as their impacts can be modeled properly. For example, in \cite{2025ApJ...989..185Z}, 
we develop our peak model to include the IA effects, in which $K_{\rm H}$ is mainly affected by the satellite IAs and both the satellite and central IAs change the moments of the shape noise field. 
For the baryonic effects, they alter both $K_{\rm H}$ and $K_{\rm{LSS}}$, and we will come back to this later.  

We have applied the peak height statistics to different survey data for cosmological constraints \citep{2015MNRAS.450.2888L, 2016PhRvL.117e1101L, 2018MNRAS.474.1116S, 2023MNRAS.519..594L}.
   
Here using observational data, we aim to demonstrate for the first time the applicability of the peak steepness statistics and in particular its advantage in probing the halo density profile. 

For that, theoretically we consider the baryonic effects following the same approach as that of the HMcode2016 \citep{2015MNRAS.454.1958M, 2016MNRAS.459.1468M}, where they are described phenomenologically 
as to reduce the concentration parameter of halos. 

Specifically, for computing the power spectrum of $K_{\rm{LSS}}$, we use HMcode2016 but with the one-halo term integrated up to $M_*$. The Sheth-Tormen halo mass function \citep{1999MNRAS.308..119S} is adopted. 
We assume the Navarro-Frenk-White (NFW) profile \citep{1996ApJ...462..563N, 1997ApJ...490..493N} for halos with the mass-concentration (M-c) relation in the form of \citep[]{2001MNRAS.321..559B} 
\begin{equation}\label{eq:m-c}
  c(M,z)=A\frac{1+z_{\rm f}}{1+z},
\end{equation}
where $A$ is the amplitude parameter, and $z_{\rm f}$ is the formation redshift of the considered halo defined with a fraction of $f=0.01$ of the final mass having collapsed.   
In fitting the nonlinear matter power spectrum, an artificial halo bloating parameter $\eta=\eta_0-0.3\sigma_8(z)$ is introduced in HMcode2016 that alters the halo profile in a mass-dependent way, where
$\sigma_8(z)$ is the rms of the linear density perturbations at redshift $z$ under the top-hat filtering of $8h^{-1}\hbox{ Mpc}$. An approximate degeneracy relation of 
$\eta_0=1.03-0.11A$ is found \citep{2015MNRAS.454.1958M}, which is used in our model calculations. 
It is shown that for the non-linear matter power spectrum, $A\approx 3.13$ is a good fit to dark-matter only simulations. For simulations with baryonic feedback, the fitted $A$ 
is systematically lower as the feedback effects get stronger \citep{2015MNRAS.454.1958M}.    
  
We need to point out that in calculating the power spectrum of $K_{\rm{LSS}}$, we keep the full two-halo term that contains the correlations between all the halos including the
contributions between the considered massive halos and the others \citep{2018ApJ...857..112Y}. In other words, for the power spectrum of $K_{\rm{LSS}}$, we do include the correlations between $K_{\rm H}$ and $K_{\rm{LSS}}$
and only exclude the $K_{\rm H}$ contributions in the one-halo term calculations. On the other hand, under the Gaussian approximation for $K_{\rm{LSS}}$, the non-Gaussian correlations between $K_{\rm H}$ and $K_{\rm{LSS}}$
are not taken into account in our modeling. The model performance has been extensively tested in, e.g., \cite{2018ApJ...857..112Y} and \cite{2023MNRAS.520.6382L}, showing well its validity for high peaks.

It is noted that in HMCode2016, an artificial bloating factor is introduced in the halo profile in order to reduce the small-scale power spectrum to be consistent with
that from simulations \citep{2015MNRAS.454.1958M, 2016MNRAS.459.1468M}, which is also needed in the dark-matter only case without baryonic feedback. That is to say, the effect of the baryonic feedback is not
reflected in the bloating factor but is characterized by the parameter $A$ in the M-c relation in HMCode2016.

For the $K_{\rm H}$ part, we employ the same halo mass function and the M-c relation as above in the calculations.
Our simulation tests show that the artificial bloating factor mentioned above for improving the power spectrum modeling should not be
included in the halo profile for the $K_{\rm H}$ calculations in our peak model. The baryonic effects are reflected by the parameter $A$ of the M-c relation,
which impacts both $K_{\rm H}$ from massive halos and the power spectrum of $K_{\rm{LSS}}$
from other halos. We will validate the model using mock data in the next section.  

We mention that here we employ HMCode2016 for calculating the power spectrum of $K_{\rm{LSS}}$, in which the influence of the baryonic feedback is described by the parameter $A$.
This framework is well in line with our peak model and can be straightforwardly implemented into the model calculations. As shown in the next section, the model works well within the statistical uncertainties.
The same approach was also adopted in \cite{2023PhRvD.108l3518L} for the HSC Y3 3$\times$2pt analyses. On the other hand, HMCode2020 incorporates more sophisticated descriptions about the baryonic effects
considering dark matter, gas and stellar components in halos and the gas expulsion with several physically motivated parameters, closely related to e.g., $\log_{10}(T_{\rm {AGN}})$ for the strength of the feedback from
active galactic nuclei (AGN) \citep{2021MNRAS.502.1401M}. Our next step aims to update our peak model incorporating HMCode2020 for the power spectrum calculation
for $K_{\rm{LSS}}$ and to consistently modify the lensing signal calculation from $K_{\rm{H}}$ taking into account multiple components.

For cosmological analyses, we consider three free parameters $(\Omega_{\rm m}, \sigma_8, A)$ with the priors shown in Table \ref{tab:priors}. 

\begin{table}
\centering
\caption{Priors.}
\label{tab:priors}
$\begin{array}{cc}
\hline \hline
\text{Parameters} & \text{Flat prior ranges}  \\ \hline {\Omega_{\mathrm{m}}} & {[0.05,0.95]}  \\  {\sigma_{8}} & {[0.20,1.60]}  \\ {A} & {[0.00,10.00]}  \\
\hline \hline
\end{array}$
\end{table}

\section{Results from mock data}\label{sec:Mock validation}

\begin{figure*}
\centering
\includegraphics[width=0.9\columnwidth]{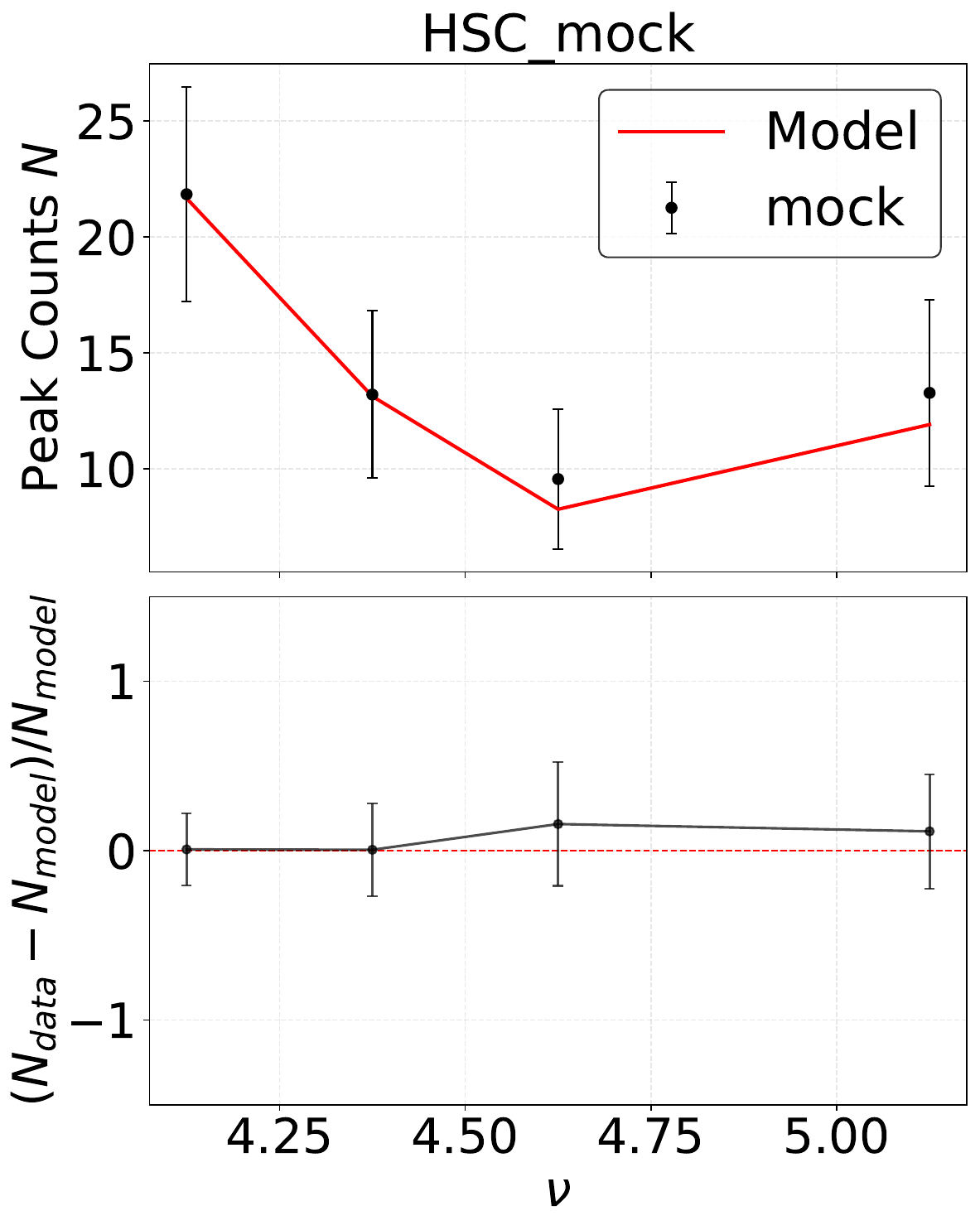}
\includegraphics[width=0.9\columnwidth]{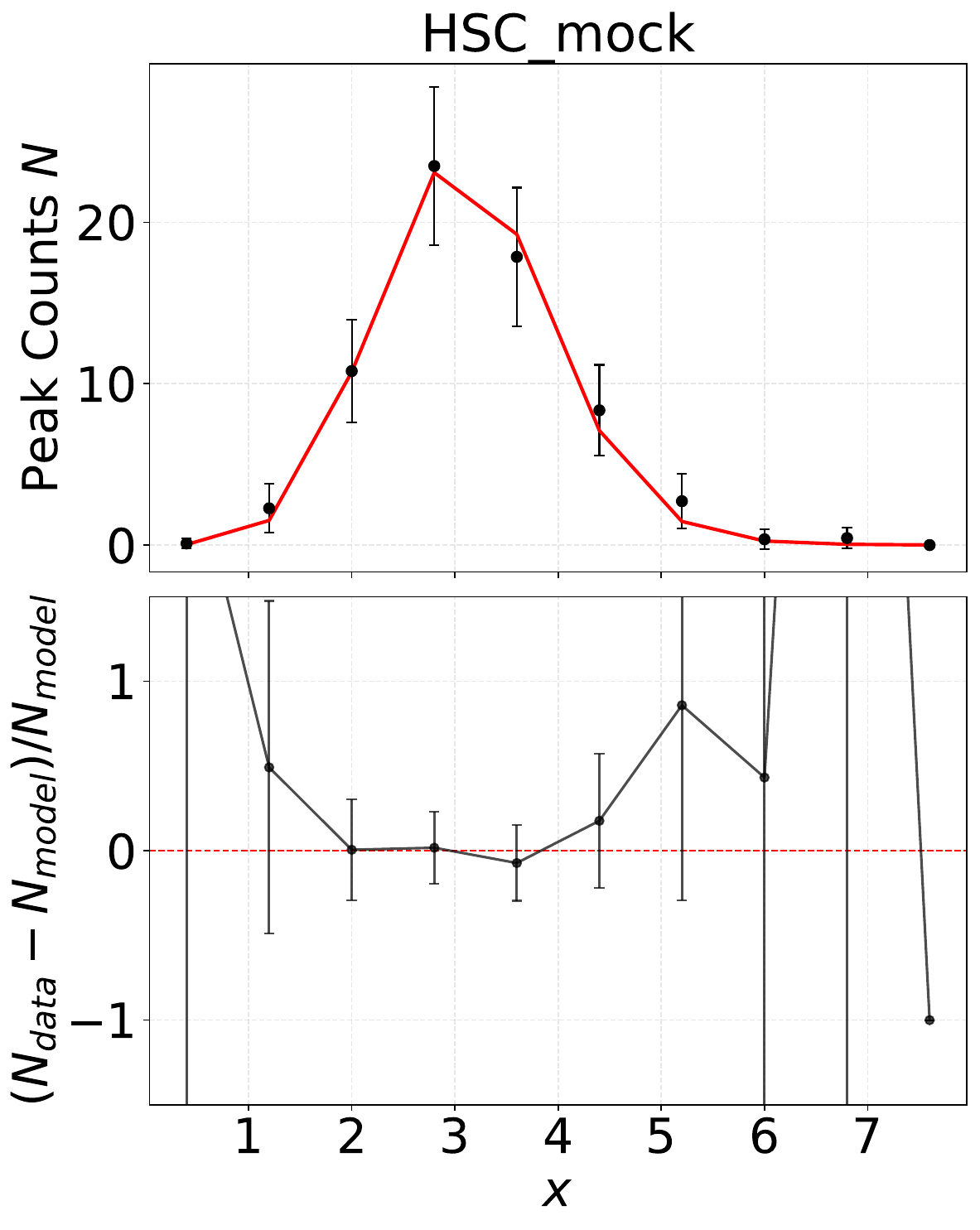}
\centering
\includegraphics[width=1.3\columnwidth]{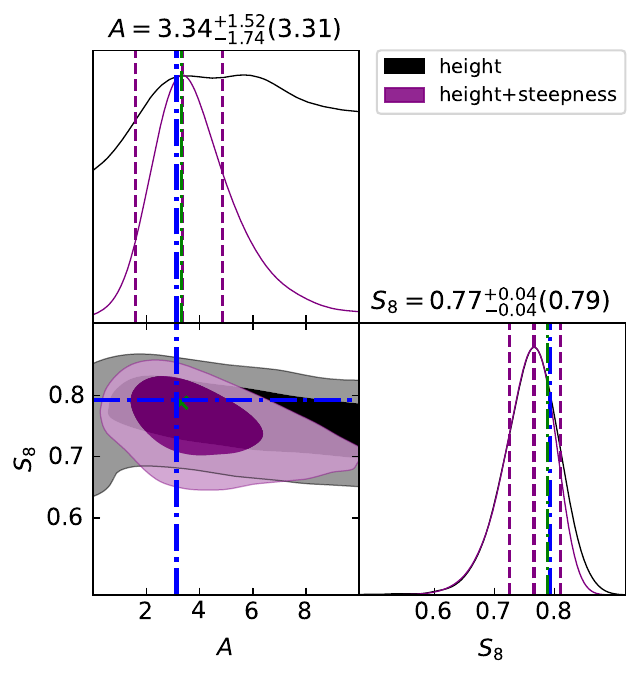}
\caption{\label{fig:HSC_mock} Upper: The average peak counts with $1\sigma$ error bars from our $1000$ HSC mocks with respect to the height $\nu$ (left) and the steepness $x$ (right).
The red lines are the model predictions and their relative differences to the mock data are presented in the bottom parts. 
Lower: The constraints from the mock data of peak height (black) and the joint height and steepness (purple) statistics.  
The 2-D contours are $68\%$ and $95\%$ CI. The blue dash-dotted lines indicate the value of $S_8$ from the input $(\Omega_{\rm m},\sigma_8)$
and $A=3.13$ from HMcode16 for the dark matter only case. The specific constraints listed are from the joint peak analyses
with the purple dashed lines showing the 1D marginalized mode point and $\pm 34\%$ CI boundaries. The green cross and dashed lines refer to the corresponding MAP values shown in the parentheses.
}
\end{figure*}

The upper panel of Figure \ref{fig:HSC_mock} shows the height (left) and steepness (right) distributions for peaks with $\nu\ge 4$, where
the data points and the error bars are the average peak counts and $1\sigma$ range for different bins from $1000$ mocks, and the red lines are our model predictions with the fiducial cosmological parameters
and $A=3.13$. For model calculations, the parameter $M_*$ should be in cluster scale. In this study, we 
take $M_*=10^{14.1}h^{-1}\hbox{M}_\odot$ that gives rise to the results in good agreement with the simulation data. The second row shows the relative differences between the data and the model predictions. Within the error bars, the two agree well. In Appendix \ref{app:Mstartest}, we present the tests on the specific choice of $M_*$ value. Within the statistical uncertainties in our analyses, the results are not very sensitive to $M_*$. We note that while $M_*$ should be physically associated with clusters at the order of $10^{14}h^{-1}\hbox{M}_\odot$, its specific value can be considered as a model parameter. In future studies with a large number of high peaks, we can fit $M_*$ simultaneously with cosmological parameters to mitigate the possible impacts of a priori choice of a fixed $M_*$ value.

We then perform Markov Chain Monte Carlo (MCMC) fittings for $(\Omega_{\rm m}, \sigma_8, A)$ based on the mock data using COSMOMC code \citep[]{2002PhRvD..66j3511L}. 
For the combined analyses of both peak height and steepness, we compute the joint covariance accordingly from the $1000$ mocks. To avoid ill-calculated inverse covariance, 
the chosen peak counts in the MCMC fitting are the bins with the lower and upper boundaries of $\nu=[4,4.25), [4.25, 4.5), [4.5, 4.75), [4.75, 5.5]$, and 
$x=[1.6,2.4), [2.4, 3.2), [3.2,4.0), [4.0,4.8]$, for the peak height and steepness respectively.
A Gaussian likelihood is adopted during fitting process. The covariance matrix and discussions about the likelihood are presented in Appendix \ref{app:likelihood}.

The results are in the lower panel of Figure \ref{fig:HSC_mock} with $S_8=\sigma_8\sqrt{\Omega_{\rm m}/0.3}$ and the prior boundary effects considered.  
The black and purple ones are from the $4$ and $4+4$ data points of the peak height only and the joint height and steepness statistics, respectively.  
The listed numbers are the marginalized 1D mode$\pm34\%$ credible interval (CI) from the joint constraints (purple lines).
The corresponding maximum-a-posteriori (MAP) values are in the parentheses and indicated by the green cross and the green lines. 
The blue lines represent the input $S_8=0.792$ and $A=3.13$ for the dark matter only case.
The constraints are in excellent agreement with the fiducial values, validating our theoretical model and the analyzing pipeline including the convergence reconstruction and the boundary/mask controls.

Clearly, the inclusion of the peak steepness statistics leads to a much better constraint on the concentration parameter $A$ than that using only the peak height statistics. 
This demonstrates strongly the value of the WL peak steepness statistics to probe the density profile of halos that serves as an important indicator for astrophysical process 
\citep[e.g.,][]{2014JCAP...04..028F,2014JCAP...08..028F,2015MNRAS.454.1958M,2020MNRAS.495.4800A}, and/or dark matter properties \citep[e.g.,][]{2010ARA&A..48..495F,2025PDU....4901965D} on the cluster scale.

\section{Potential systematics}\label{sec:systematics}
We note that the mocks retain observed galaxy distributions, and thus the boundary and mask effects are included. On the other hand, the mocks cannot properly account for all the
potential systematics in the real observational data. For example, because the structures in the simulations do not correlate with the structures
in the observed galaxy distributions, the effects from the cluster member contamination to be described in Sec.~\ref{sec:cluster member contaminations} are not shown up in mocks \citep[e.g.,][]{2023MNRAS.519..594L}. 
In this section, we discuss potential systematics on our analyses, including how to model the effects of cluster member contaminations by performing specific single-halo simulations, 
the possible impact of galaxy intrinsic alignments (IA) using a simulation with semi-analytical galaxy formation
\citep{2022ApJ...940...96Z, 2018ApJ...853...25W}, and the B-mode tests. As shown in \cite{2023MNRAS.519..594L}, the errors of the shear and photo-z measurements of S16A have negligible effects comparing to the 
statistical uncertainties.

\subsection{Modeling the cluster member contaminations}\label{sec:cluster member contaminations}
For a shear galaxy sample with a wide redshift distribution, it includes inevitably galaxies that are satellite members of foreground massive clusters.
These galaxies do not carry the WL signals
arising from their host clusters, and can lead to a dilution effect in estimating the WL signals in the cluster regions. This can be understood as follows.
The average WL shear signal from a cluster can be schematically written as
$\langle{\bm{\gamma}}\rangle_{\rm{cl}}=\sum_i^{N_{\rm{gal}}}{\bm{\gamma}}/N_{\rm {gal}}$, where the sum is over all the source galaxies in the considered region.
The member galaxies have no $\bm{\gamma}$ from their host and thus do not contribute to the nominator, but contribute to the denominator, causing a systematically lower estimation for the cluster WL signal.
In cluster lensing studies, such member contamination effects are normally controlled by selecting source galaxies with their redshifts above the known cluster redshift.
This is not applicable in WL peak analyses, and thus the impacts from the cluster member contamination need to be investigated carefully. Besides the dilution effect,
the contamination also changes the shape noise level in cluster regions that also matters in our peak studies.

In our previous studies \citep{2018MNRAS.474.1116S}, we develop and validate a methodology to include these impacts into our theoretical model using the member contamination information estimated from the cluster catalogs in the
survey fields. The dilution effects on $K_{\rm H}$ are measured through single-halo simulations for different mass and redshift bins, and the impacts on the noise field are taken into account by considering halo regions and 
field regions separately. This same approach is adopted in \cite{2023MNRAS.519..594L} for HSC Y1 tomographic studies of peak height statistics. 

\begin{figure*}
\centering
\includegraphics[width=2.0\columnwidth]{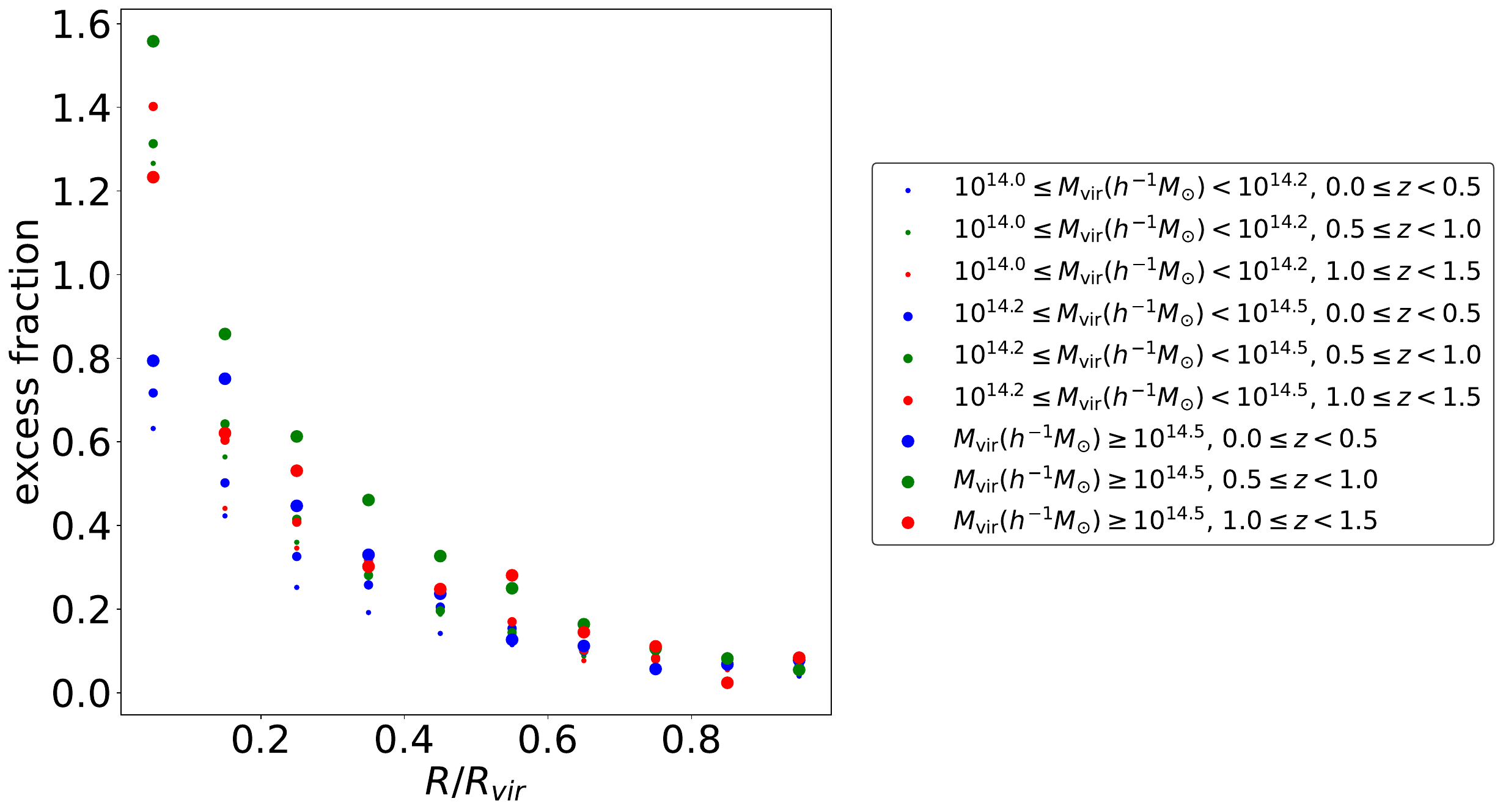}
\caption{\label{fig:excess} 
The member excess fractions for clusters with different mass and redshift bins, measured from the WZL catalog.   
}
\end{figure*}

Because the peak steepness is more sensitive to the spatial distribution of $K_{\rm H}$, here we extend this treatment to consider the dilution effects on the profiles of $K_{\rm H}$ and their derivatives 
rather than estimating a single dilution factor as that in \cite{2018MNRAS.474.1116S, 2023MNRAS.519..594L}. We also include the radial dependence of the shape noise in massive halos in the model calculations.  
For the contamination information, similar to that in \cite{2023MNRAS.519..594L}, we measure the member excess in clusters of different mass and redshift bins using the cluster catalog of \cite{2021MNRAS.500.1003W} (WZL catalog).
The results are shown in Figure \ref{fig:excess}. 

For single-halo simulations, we place a lens NFW halo with the considered mass and redshift in the central region of an area of $1.2\times 1.2\deg^2$. Galaxies with the same redshift distribution, the number density
and the intrinsic ellipticity distribution as that of the observed sample are first randomly populated in this area and their shear signals due to the central halo are calculated. 
We then rearrange them to mimic the observed member contamination profile by randomly moving galaxies outside the halo region into it. 
The shear signals of those moved-in galaxies are eliminated. The convergence reconstructions are done respectively from the galaxy samples before and after   
rearrangement. For each halo case, we do $1000$ single-halo simulations by moving different galaxies from outside to inside of the halo region, and obtain the average profiles. 

\begin{figure*}
\centering
\includegraphics[width=2.0\columnwidth]{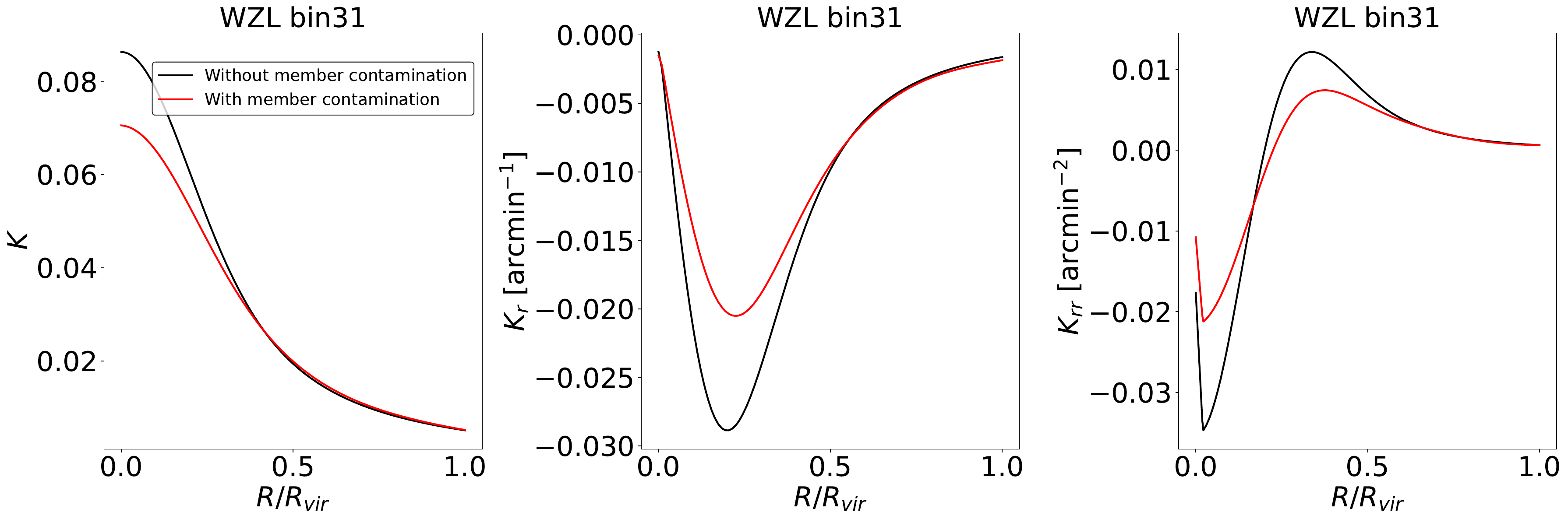}\\
\hspace{0.1cm}
\includegraphics[width=0.95\columnwidth]{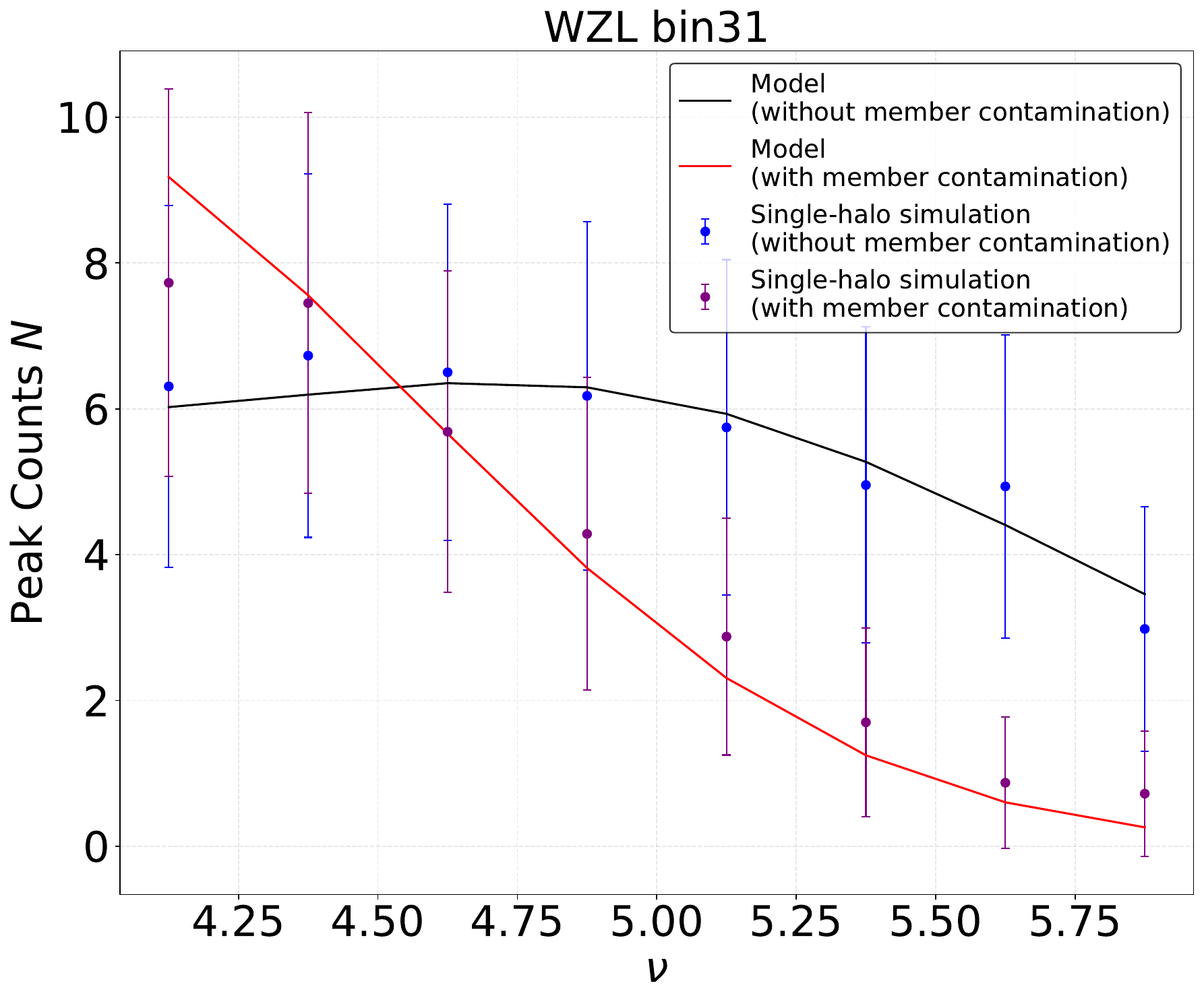}
\hspace{0.3cm}
\includegraphics[width=0.95\columnwidth]{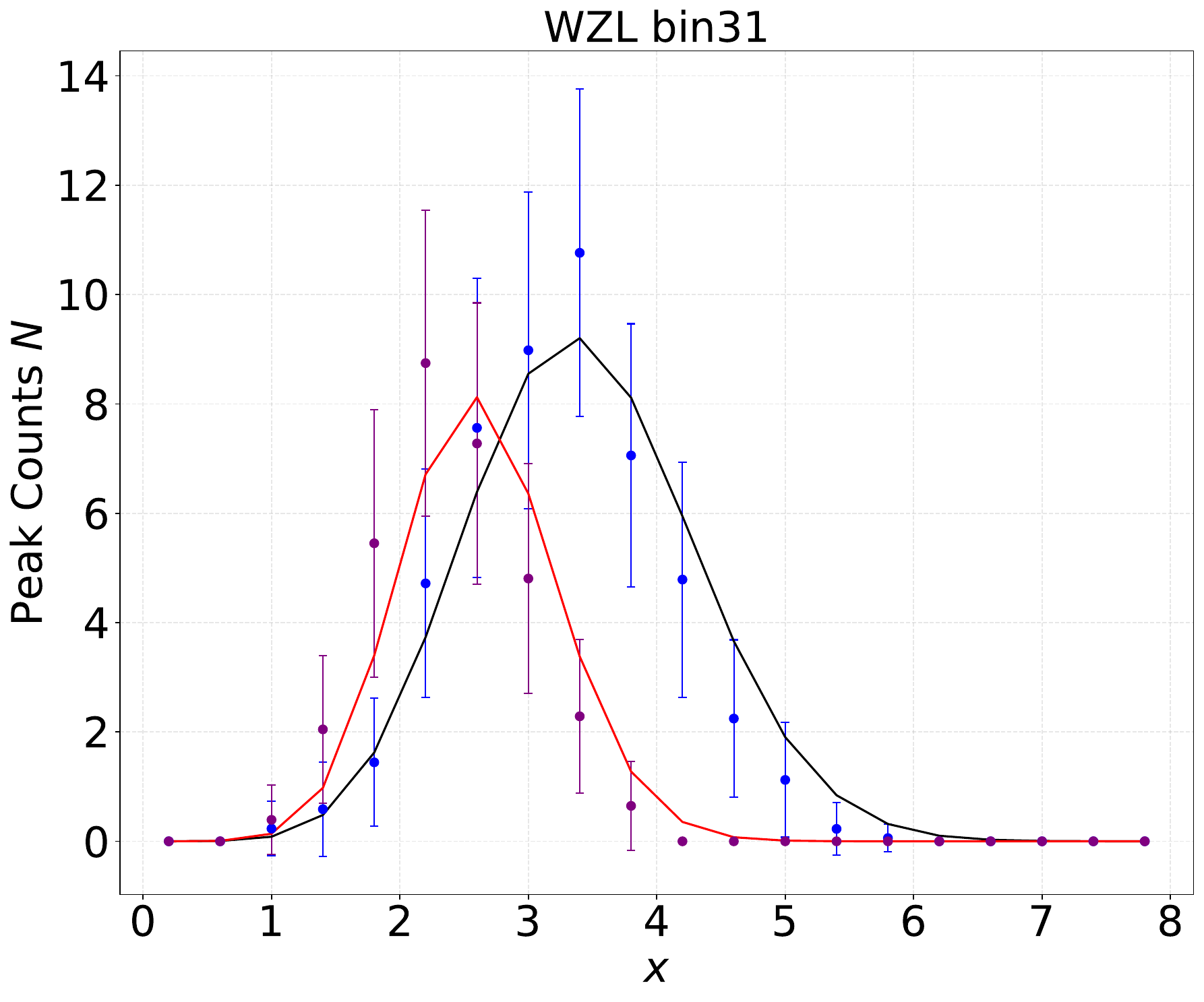}
\caption{\label{fig:boost} 
Upper: The profile changes of $K_{\rm H}$ (left) and its first (middle) and second (right) derivatives due to member contaminations for a cluster with $M=10^{14.60}h^{-1}M_\odot$ at redshift $z=0.3$. 
Lower: Comparison of simulated data and the model predictions with the left and right for peak height and steepness statistics, respectively. 
Peaks with $\nu\ge 4$ are considered. 
}
\end{figure*}

In Figure \ref{fig:boost}, we show the results for the case of halo mass $M=10^{14.60}h^{-1}M_\odot$ at redshift $z=0.3$.
The upper panels show the profiles without (black) and with (red) member contaminations for the halo $K_{\rm H}$ (left) and its first (middle) and second (right) derivatives.
The lower panels show the corresponding comparisons of our model predictions for the peak height (left) and steepness (right) statistics for peaks with $\nu\ge 4$ with the data from the $1000$ single-halo simulations.
The peak numbers and error bars are for the case with the area of $60\deg^2$, similar to the observed area we used.

Our model performs well to account for the effects of member contaminations. 

In the overall model calculations considering halos with $M\ge M_*$, we take into account the dilution profile changes for different mass and redshift bins and the noise changes from 
the corresponding member clustering.    

\subsection{The IA effects}
In weak lensing studies, galaxy IAs are a major source of systematics \citep[e.g.,][]{2015PhR...558....1T,2025A&ARv..33....5C}. For peak analyses, they affect the noise properties \citep[e.g.,][]{2007ApJ...669...10F,2025ApJ...989..185Z}. 
Additionally, the satellite IAs can further change the lensing signal of a cluster besides that from member contaminations discussed above \citep[e.g.,][]{2022MNRAS.511.2075Z,2022ApJ...940...96Z}. 

To evaluate the potential IA impacts, we use the simulated data from \cite{2022ApJ...940...96Z} based on a large ray-tracing simulation with semi-analytic galaxy formation that includes the galaxy clustering 
naturally \citep{2018ApJ...853...25W}. The IAs of central galaxies
follow their host halo orientations for ellipticals and are perpendicular to the spin of host halos for disk galaxies. For satellites, different IA settings are analyzed in \cite{2022ApJ...940...96Z}. Here we consider
satellite IAs that are averagely radially aligned toward the center of their host cluster in 3D with a dispersion $\sigma_{\theta}$. The smaller the $\sigma_{\theta}$ is, the stronger the IA effects are. 
Given that no strong satellite IAs have been detected, we take $\sigma_{\theta}=60^{\circ}$ here. With the IAs being set, we construct galaxy samples from the simulation following the redshift distribution and the number density
of the HSC observational data, and further perform the convergence reconstructions and the peak analyses. We also generate similar galaxy samples by first rotating them randomly to remove IAs and then
adding the shear signals to construct a reference case without IAs for comparison. 

\begin{figure*}
\centering
\includegraphics[width=0.9\columnwidth]{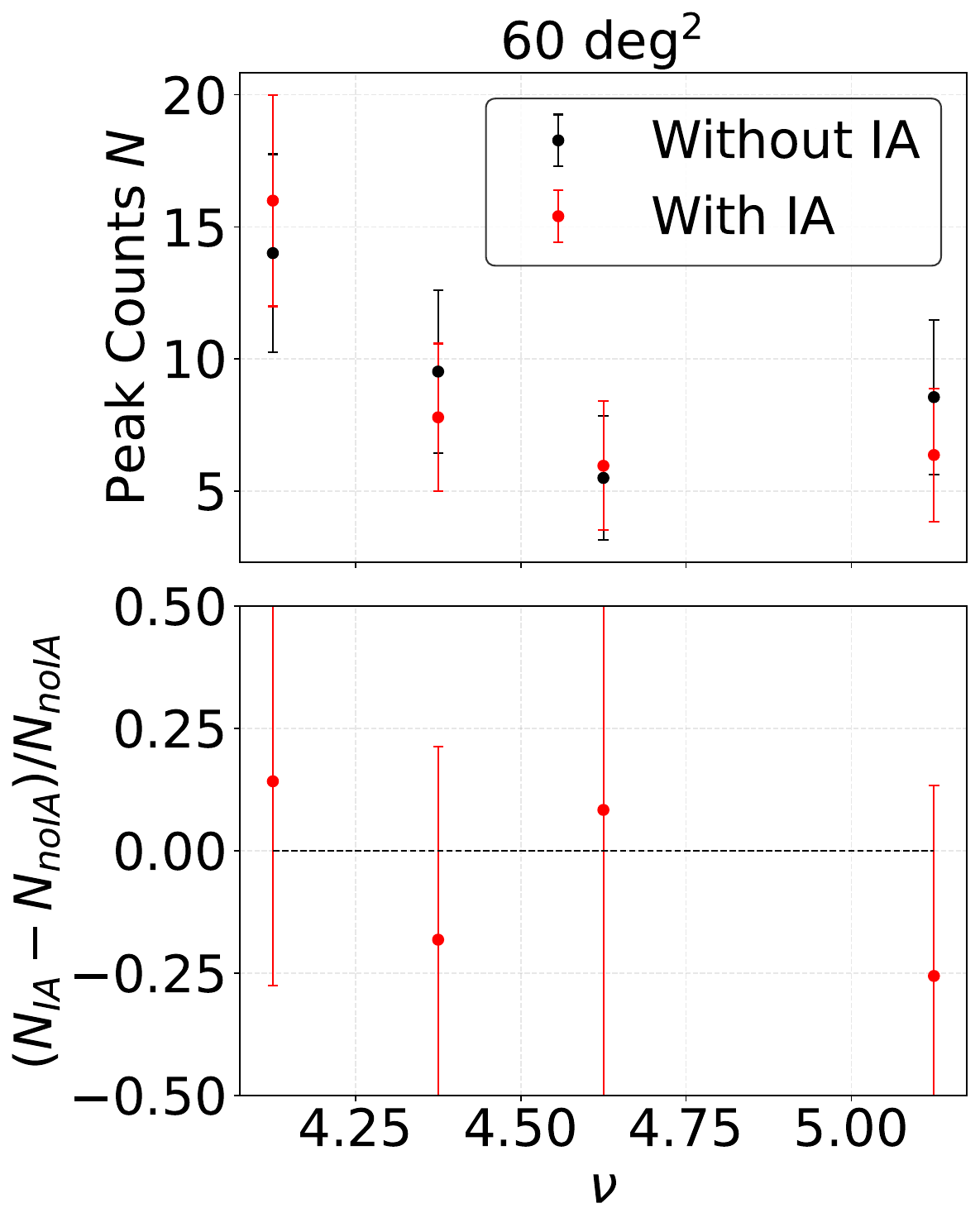}
\includegraphics[width=0.9\columnwidth]{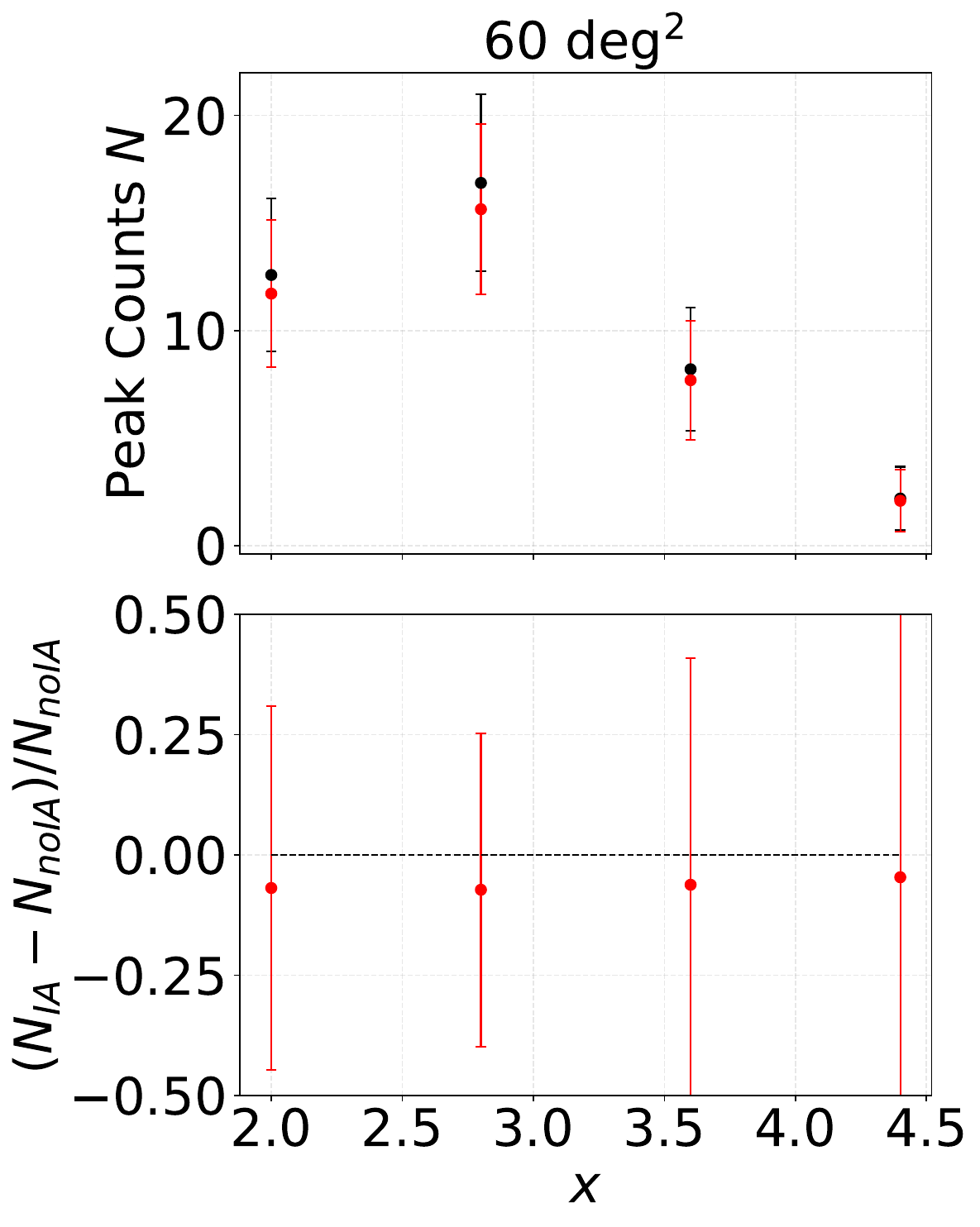}
\caption{\label{fig:IA} The peak height (left) and steepness (right) distributions without (black) and with (red)
IA effects. The lower panels show the relative differences between the two cases. The peak binning is the same as our MCMC analyses.}
\end{figure*}

Figure \ref{fig:IA} shows the peak height (left) and steepness (right) distributions with the same binning as that used in our MCMC analyses. The results without (black) and with (red) IAs are compared. 
It is seen that within the statistical error bars corresponding to the survey area of $60\deg^2$ similar to that used in our observational analyses, the IA impacts are minimal. 

\subsection{B-mode test based on peak analyses}
In the WL regime, the ideal lensing signals should contain only E-mode arising from the gravitational effects of large-scale structures. 
Higher order physical effects can generate B-mode signals, but they are typically much smaller than the E-mode counterpart
\citep[]{2010A&A...523A..28K, 2002ApJ...568...20C,2013MNRAS.435..194C, 2002A&A...389..729S, 2003ApJ...583...58W,2009ApJ...702..593S}. Therefore in WL analyses, B-mode tests are normally 
used to identify possible problems in data processing and shear measurements.    

For the HSC S16A shear sample, in \cite[]{2020PASJ...72...16H}, B-mode tests are done in terms of cosmic shear 2pt correlations, and no significant B-mode signals are detected.
Here we present our B-mode test from the view of peak distributions. In the ideal WL case, a reconstructed convergence field has only the real part \citep[e.g.,][]{2001PhR...340..291B}. Considering the intrinsic shapes of galaxies, 
the imaginary part of a reconstructed field should be consistent with the pure shape noise field, and strong deviations hint to the possible existence of B-mode.
Here we analyze the imaginary parts of the reconstructed fields from both the observational data and from the simulation mocks as well as from the observed shape noise field with randomly rotated galaxies. 

\begin{figure*}
\centering
\includegraphics[width=0.9\columnwidth]{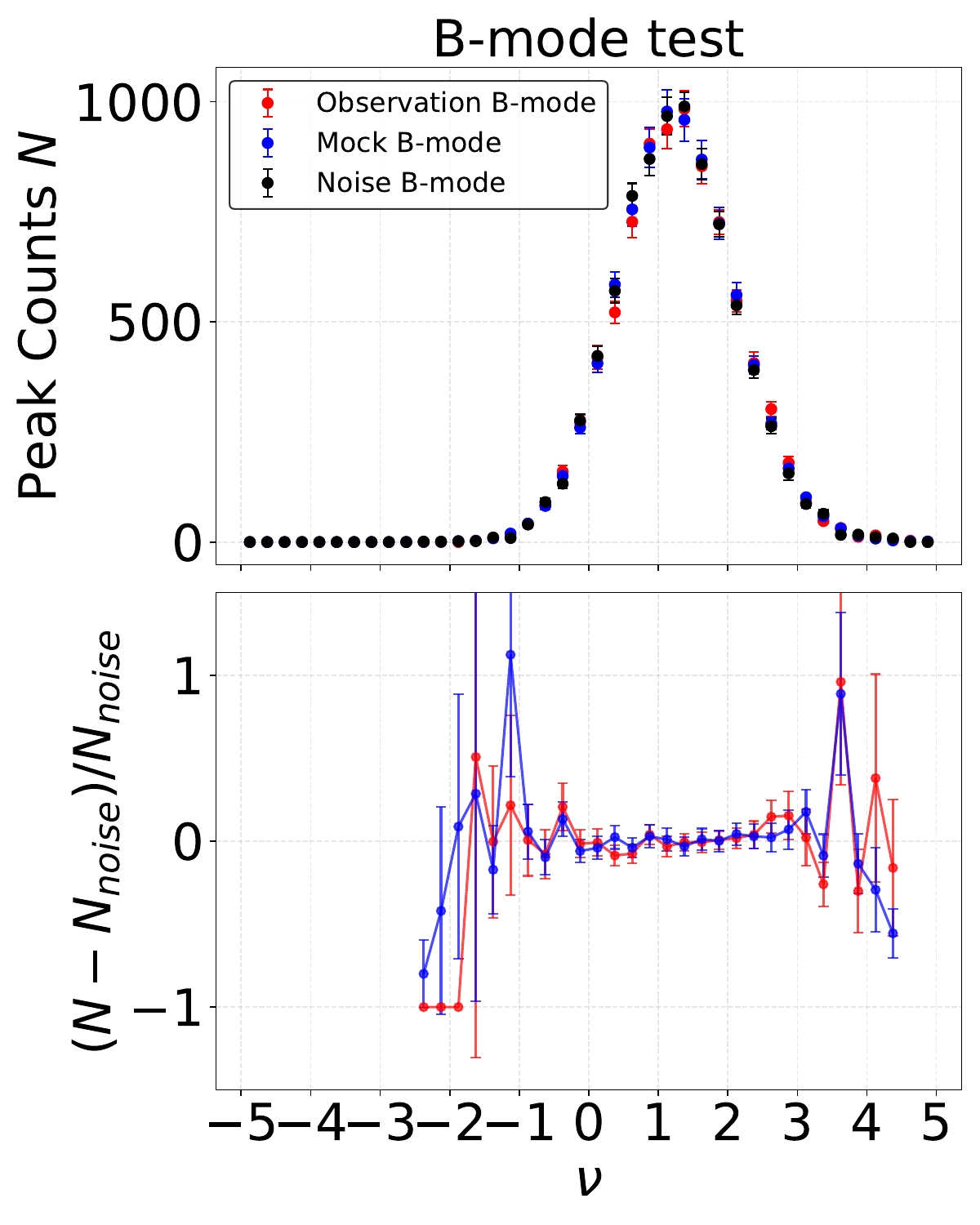}
\includegraphics[width=0.9\columnwidth]{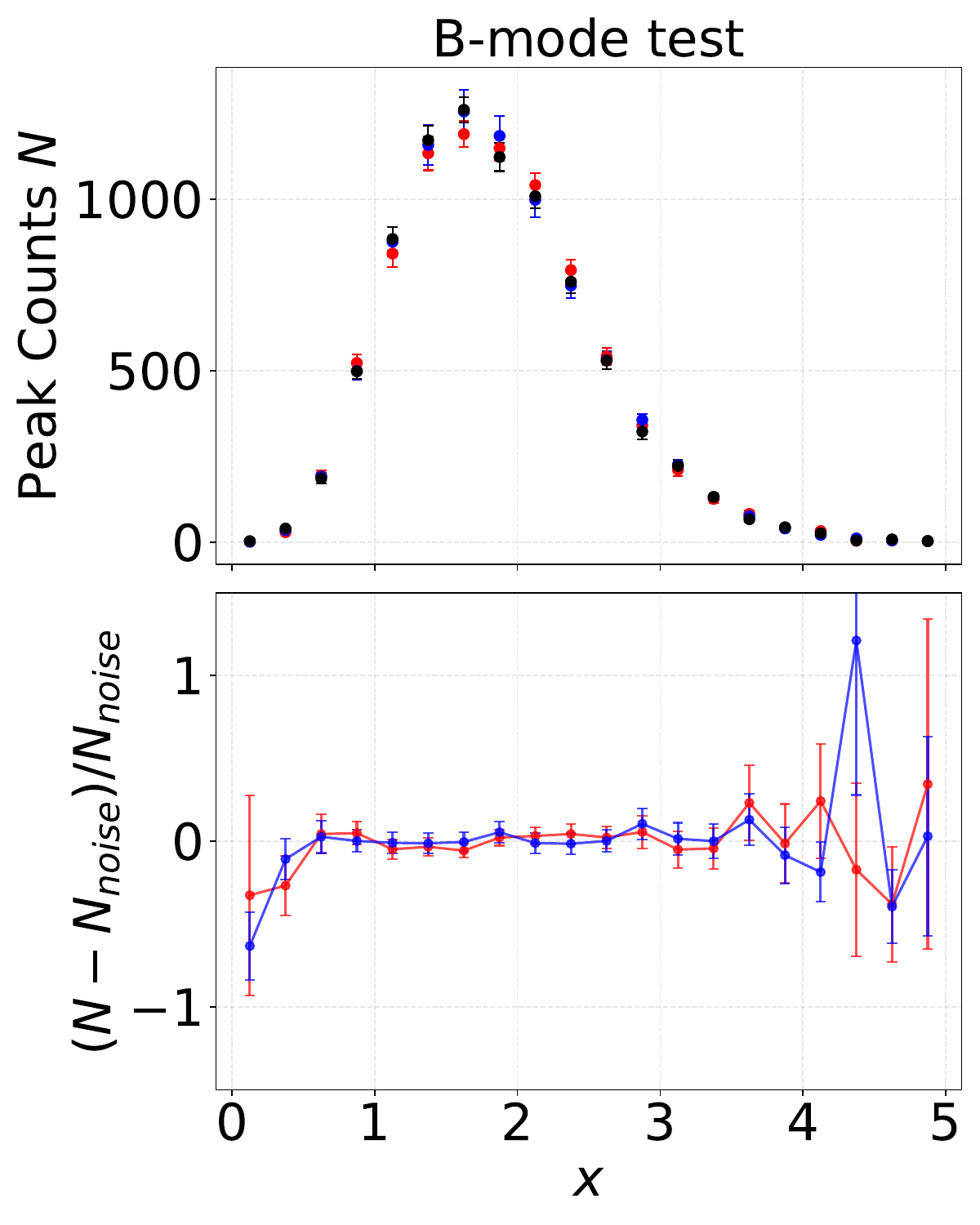}
\caption{\label{fig:B-mode} The peak distributions from HSC mock (blue), observational (red) and pure shape noise (black) B-mode maps with respect to the height $\nu$ (left) and the steepness $x$ (right). 
The lower panels show the relative differences with respect to the black ones. 
}
\end{figure*}

Figure \ref{fig:B-mode} shows the peak height (left) and steepness (right) distributions with the red and blue data points for mock (averaged over the 1000 mocks) and observational results, respectively. 
The black points are the results from the shape noise fields. The lower panels show the corresponding relative differences with respect to black ones. 
Within the error bars, both the mock results and the observational results are consistent with the case of the shape noise fields, showing negligible B-mode effects.

\section{Results from observational data}\label{sec:Observational results}
With the mock validation and the discussions on potential systematics, we apply the analyses to the observational data. Here we employ the revised model that includes the effects from the cluster member contaminations on the profiles of $K_{\rm H}$ and its derivatives and on the noise changes 
as demonstrated in section \ref{sec:cluster member contaminations}.

\begin{figure*}
\centering
\includegraphics[width=1.3\columnwidth]{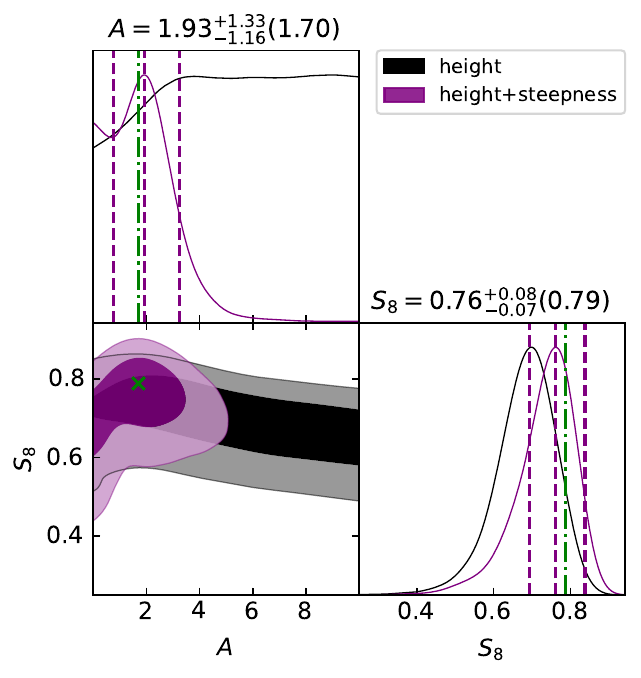}\\
\includegraphics[width=0.9\columnwidth]{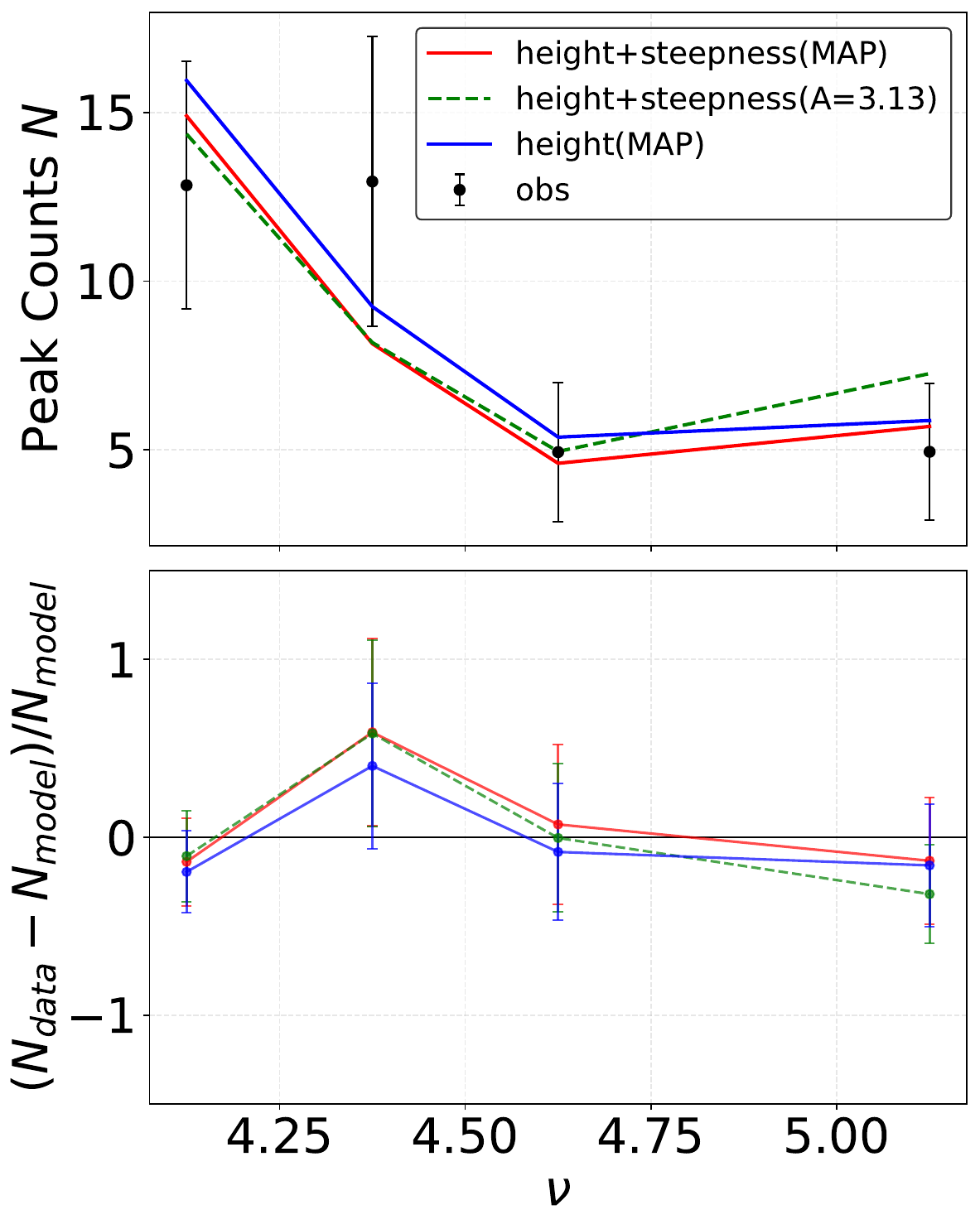}
\includegraphics[width=0.9\columnwidth]{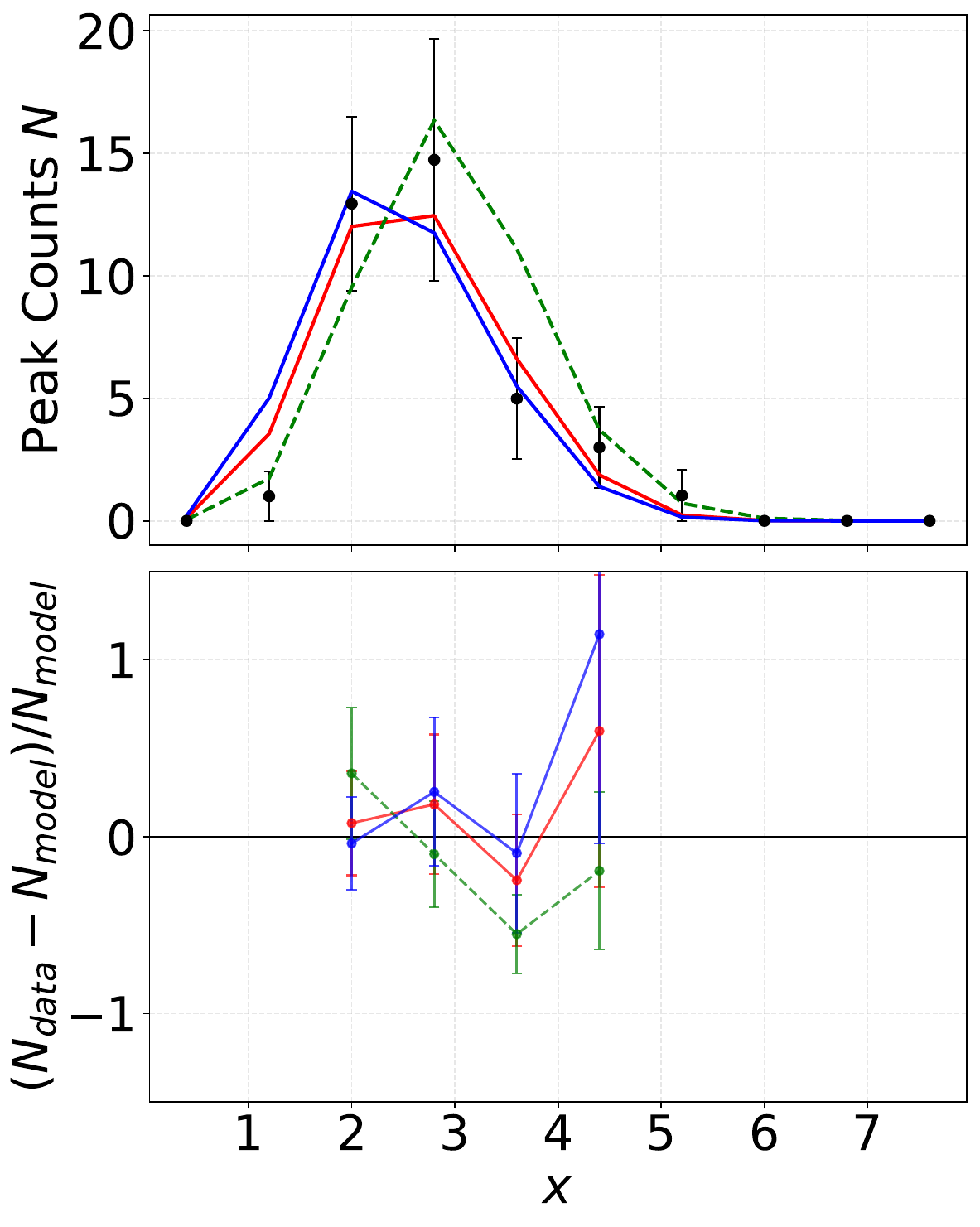}
\caption{\label{fig:HSC_obs_MCMC} Upper: The constraints from observed peak counts with the meaning of lines and symbols the same as those in the lower panel of Figure \ref{fig:HSC_mock}. 
Lower: The observed data and the model results with the MAP parameters fitted from peak height only (blue) and from the combined height+steepness analyses (red). The green dashed lines
are the model predictions with the cosmological parameters the same as the red line but the concentration parameter $A=3.13$. Note that $4$ peak height and $4$ steepness data are used in the fitting, and their
relative differences with different model results are presented in the bottom parts. 
}
\end{figure*}

The MCMC fitting results are presented in the upper panel of Figure \ref{fig:HSC_obs_MCMC}. 
The derived constraint is $S_8=0.76^{+0.08}_{-0.07}(\hbox{MAP}=0.79)$ from the joint peak analyses, which is consistent with other studies using HSC WL data \citep[e.g.,][]{2023MNRAS.519..594L,2023PhRvD.108l3518L,2025arXiv250814019S}.
For the parameter $A$, it is $A=1.93^{+1.33}_{-1.16}(\hbox{MAP}=1.70)$, about $1\sigma$ lower than that from the dark matter only simulated mocks. The lower panels present the observed data and the model calculations with 
the fitted MAP parameters from peak height only (blue) and from the combined height+steepness analyses (red), respectively. For comparison, we also plot the model predictions with the cosmological parameters the same as the 
red line but with $A=3.13$ (green line). The green line deviates from the data more in the steepness statistics than in the peak height case due to their different sensitivities to $A$. 

This low $A$ is more in line with the fitted value in HMcode2016 \citep[]{2015MNRAS.454.1958M} for the power spectrum from OverWhelmingly Large Simulations (OWLS) with AGN feedback \citep[]{2010MNRAS.402.1536S,2011MNRAS.415.3649V}.
It is also consistent with the constraint derived from HSC Y3 cosmic shear 2pt analyses noting a very tight prior of $(2,3.13)$ applied to $A$ in their studies \citep{2023PhRvD.108l3518L}.
In \cite{2023MNRAS.518.5340C} and \cite{2023A&A...678A.109A}, small-scale cosmic shear 2pt from Dark Energy Survey (DES) Year-3 data are analyzed with the baryon correction model (BCM) \citep{2020MNRAS.495.4800A}
in which the baryonic effects are explicitly modeled by different components including the central galaxy, the hot gas bound to a halo, the ejected gas out of a halo, and the dark matter part. 
Their results also point to strong baryonic feedback on cluster scales, consistent with the OWLS-like simulations with strong AGN feedback. Studies utilizing other probes have reached similar conclusions 
\citep[e.g.,][]{2025PhRvD.112h3509H, 2025arXiv250910455S}.
  
Comparing to WL cosmic shear 2pt and peak height statistics, the peak steepness statistics is more sensitive to the total density profile of massive halos. 
Also in principle it is free of selection effects existed in e.g., X-ray cluster studies \citep[e.g.,][]{2025arXiv250910455S}. 
Our analyses here demonstrate well its potential to probe physical processes that can alter the matter distribution of halos, such as baryonic effects and dark matter properties, in addition to constrain cosmological parameters. 

\section{Summary and Discussion}\label{sec:Summary and Discussion}
We perform the first-time application of the new WL peak steepness statistics to observational data with the 
main aim to demonstrate its strong potential in cosmological studies.  
Our analyses validated by simulation mocks show well its sensitivity to the density profile of halos, and 
the observational constraint results in a low concentration parameter $A$ comparing to that from dark-matter-only mocks, favoring strong baryonic feedback at the cluster scale.     

For potential systematics, we include the effects from member contaminations explicitly in our model. 
We also analyze the IA effects and the B-mode, and conclude that they are insignificant.  

Having seen the signs of strong feedback, more accurate modeling about the baryonic effects on the density distribution of halos than the one we adopt here is desired. As a next step, we will incorporate BCM-like descriptions into
our theoretical peak model, which can allow us to derive information that is more physically linked to different astrophysical processes.
Apart from probing baryonic effects, the peak steepness statistics also has a promising potential to constrain different dark matter models through their influence on halo density profile, an important aspect that deserves future investigations.

\begin{acknowledgments}
We are grateful for the discussions with Amol Ravindra Upadhye and Shiyan Zhong. This study is supported in part by the NSFC grant 11933002 and the grant from the China Manned Space Projects with No. CMS-CSST-2021-A01 and CMS-CSST-2025-A05. 
Z.H.F. also acknowledges the support from NSFC grant U1931210.  X.K.L. acknowledges the support from NSFC grant No. 12173033,  National Key R\&D Program of China No. 2022YFF0503403, the grants from the China Manned Space Projects with No.  CMS CSST-2025-A03, CMS-CSST-2025-A05, and CMS-CSST-2021-B01, the “Yunnan Key Laboratory of Survey Science”with project No. 202449CE340002 and a “Yunnan Provincial Top Team Projects” with project No. 202305AT350002. S.Y. is supported by the NSFC grant (No. 12203062), the science research grants from the China Manned Space Project with No. CMS-CSST-2021-B01 and the CAS Project for Young Scientists in Basic Research (No. YSBR-092). Q.W. is supported by the NSFC grant (Nos. 12588202).

The Hyper Suprime-Cam (HSC) collaboration includes the astronomical communities of Japan and Taiwan, and Princeton University. The HSC instrumentation and software were developed by the National Astronomical Observatory of Japan (NAOJ), the Kavli Institute for the Physics and Mathematics of the Universe (Kavli IPMU), the University of Tokyo, the High Energy Accelerator Research Organization (KEK), the Academia Sinica Institute for Astronomy and Astrophysics in Taiwan (ASIAA), and Princeton University. Funding was contributed by the FIRST program from the Japanese Cabinet Office, the Ministry of Education, Culture, Sports, Science and Technology (MEXT), the Japan Society for the Promotion of Science (JSPS), Japan Science and Technology Agency (JST), the Toray Science Foundation, NAOJ, Kavli IPMU, KEK, ASIAA, and Princeton University. 

This paper makes use of software developed for Vera C. Rubin Observatory. We thank the Rubin Observatory for making their code available as free software at http://pipelines.lsst.io/.

This paper is based on data collected at the Subaru Telescope and retrieved from the HSC data archive system, which is operated by the Subaru Telescope and Astronomy Data Center (ADC) at NAOJ. Data analysis was in part carried out with the cooperation of Center for Computational Astrophysics (CfCA), NAOJ. We are honored and grateful for the opportunity of observing the Universe from Maunakea, which has the cultural, historical and natural significance in Hawaii.
\end{acknowledgments}

\hbox{
}

\appendix
\section{Theoretical formulae for model calculations\label{app:model}}
We define $\nu_{\rm T}=\nu\sigma_{\rm N,0}/\sigma_{\rm T,0}$ and $x_{\rm T}=x\sigma_{\rm N,2}/\sigma_{\rm T,2}$, where the total moments of the joint Gaussian field $K_{\rm{LSS}}+N$
are $\sigma_{\rm T,i}^2=\sigma_{\rm{LSS},i}^2+\sigma_{\rm N,i}^2$ ($i=0,1,2$). In regions occupied by massive halos, the peak number density distribution is calculated from the Gaussian random field $K_{\rm{LSS}}+N$ modulated by 
$K_{\rm H}$ and summing over all halos with $M\ge M_*$. For field regions without halos of $M\ge M_*$, it is computed directly from the Gaussian random field of $K_{\rm{LSS}}+N$.

Specifically, the peak distributions in terms of the height $\nu_{\rm T}$ and the steepness $x_{\rm T}$ at a specific location in a massive halo region
are given respectively as follows \citep[]{2023MNRAS.520.6382L}:
\begin{equation}\label{eq:height number density}
\begin{aligned}
\hat{n}_{\text {peak }}\left(\nu_{\rm T}\right)=& \exp \left[-\frac{\left(K_{\rm H}^{1}\right)^{2}+\left(K_{\rm H}^{2}\right)^{2}}{\sigma_{\rm T,1}^{2}}\right]\left\{\frac{1}{2 \pi \theta_{\rm T*}^{2}} \frac{1}{(2 \pi)^{1 / 2}}\right\} \\
& \times \exp \left(-\frac{1}{2}u(\nu_{\rm T})^{2}\right) \\
& \times d \nu_{\rm T}\int_0^{\infty} \frac{d x_{\rm T}}{\left[2 \pi\left(1-\gamma_{\rm T}^{2}\right)\right]^{1 / 2}} \\
& \times \exp \left[-\frac{\left(m(x_{\rm T})-\gamma_{\rm T}u(\nu_{\rm T})\right)^{2}}{2\left(1-\gamma_{\rm T}^{2}\right)}\right] \times F\left(x_{\rm T}\right)
\end{aligned}
\end{equation}
and
\begin{equation}\label{eq:steepness number density}
\begin{aligned}
\hat{n}_{\text {peak }}\left(x_{\rm T}\right)\bigg{|}_{\nu_{\rm T}\ge\nu_{\text{cut}}}=& \exp \left[-\frac{\left(K_{\rm H}^{1}\right)^{2}+\left(K_{\rm H}^{2}\right)^{2}}{\sigma_{\rm T,1}^{2}}\right]\\
& \times \left\{\frac{1}{(2 \pi \theta_{\rm T*})^{2}} \frac{(2\pi)^{1/2}}{2}\right\} \\
& \times \exp \left(-\frac{1}{2}m(x_{\rm T})^{2}\right) \\
& \times \text{erfc}(t(\nu_{\text{cut}},x_{\rm T}))\\
& \times F\left(x_{\rm T}\right)d x_{\rm T},
\end{aligned}
\end{equation}
where erfc$(x)$ is the complementary error function and $\nu_{\rm {cut}}$ is a lower cut on the peak height.
For convenience, we define some combined variables and expressions as
\begin{equation}\label{eq:fx}
\begin{aligned}
& \gamma_{\rm T}=\frac{\sigma_{\rm T,1}^2}{\sigma_{\rm T,0}\sigma_{\rm T,2}},\  \theta_{\rm T*}=\frac{\sqrt{2}\sigma_{\rm T,1}}{\sigma_{\rm T,2}}, \\
& u(\nu_{\rm T}) = \nu_{\rm T}-\displaystyle{\frac{K_{\rm H}}{\sigma_{\rm T,0}}}, \\
& m(x_{\rm T}) = x_{\rm T}+\displaystyle{\frac{K_{\rm H}^{11}+K_{\rm H}^{22}}{\sigma_{\rm T,2}}}, \\
& t(\nu_{\rm T},x_{\rm T}) = \displaystyle{\frac{u(\nu_{\rm T})-\gamma_{\rm T}m(x_{\rm T})}{[2(1-\gamma_{\rm T}^2)]^{1/2}}},\\
\end{aligned}
\end{equation}
and
\begin{equation}
\begin{aligned}
F\left(x_{\rm T}\right)=& \exp \left[-\left(\frac{K_{\rm H}^{11}-K_{\rm H}^{22}}{\sigma_{\rm T,2}}\right)^{2}-4 \frac{\left(K_{\rm H}^{12}\right)^{2}}{\sigma_{\rm T,2}^{2}}\right] \\
& \times \int_{0}^{1 / 2} d e_{\rm T} \  8\left(x_{\rm T}^{2} e_{\rm T}\right) x_{\rm T}^{2}\\
& \times \left(1-4 e_{\rm T}^{2}\right) \exp \left(-4 x_{\rm T}^{2} e_{\rm T}^{2}\right) \\
& \times \int_{0}^{\pi} \frac{d \theta_{\rm T}}{\pi} \exp \left[-4 x_{\rm T} e_{\rm T} \cos 2 \theta_{\rm T}\left(\frac{K_{\rm H}^{11}-K_{\rm H}^{22}}{\sigma_{\rm T,2}}\right)\right] \\
& \times \exp \left(-8 x_{\rm T} e_{\rm T} \sin 2 \theta_{\rm T} \frac{K_{\rm H}^{12}}{\sigma_{\rm T,2}}\right).
\end{aligned}
\end{equation}

Then the overall peak number density distribution in halo regions can be written as 
\begin{equation}\label{eq:Peakhalo}
\begin{aligned}
n_{\text {peak }}^{H}(y)=& \int d z \frac{d V(z)}{d z d \Omega} \int_{M_{*}}^{\infty} d M n(M, z) \\
& \times \int_{0}^{\theta_{\text {vir }}} d \theta(2 \pi \theta) \hat{n}_{\text {peak }}^{H}(y, M, z, \theta),
\end{aligned}
\end{equation}
where $dV$ and $d\Omega$ are the cosmic volume and solid angle elements, respectively, $n(M, z)$ is the halo mass function, and ${\theta_{\text {vir }}}$ is the angular virial radius of a halo with mass $M$ at redshift $z$.
Here $y$ can be $\nu_{\rm T}$ or $x_{\rm T}$. 

For field regions, the peak number density distribution is given by 
\begin{equation}
\begin{aligned}
n_{\text {peak }}^{N}(y)=& \frac{1}{d \Omega}\times\{\hat{n}_{\text {\rm{ran} }}(y)[d \Omega- \\
&\int d z \frac{d V(z)}{d z} \int_{M_{*}}^{\infty} d M n(M, z)\left(\pi \theta_{\text {vir }}^{2}\right)]\},
\end{aligned}
\end{equation}
where $\hat{n}_{\text {\rm{ran} }}(y)$ is the mean number density of the Gaussian random field $K_{\rm{LSS}}+N$ without $K_{\rm H}$ modulations.

Finally we have the total peak number density distribution 
\begin{equation}
n_{\text {peak }}(y) =n_{\text {peak }}^{H}(y) +n_{\text {peak }}^{N}(y) .
\end{equation}

\section{Dependence on the choice of \texorpdfstring{$M_*$}{M* } value\label{app:Mstartest}}
As described above, in our halo model for WL peaks, massive halos with $M_{\rm{vir}}\geq M_*$ are considered to be dominant contributors for high peaks characterized by $K_{\rm H}$. 
Previous studies have suggested that $M_*$ should correspond to the cluster scale around $\sim 10^{14.0}h^{-1}\hbox{M}_\odot$~\citep[e.g.,][]{2018MNRAS.478.2987W,2018ApJ...857..112Y, 2025MNRAS.538..755B,2024OJAp....7E..90C,2025OJAp....8E...2C}. 
In the fiducial analyses shown in the main text, we adopt $M_*=10^{14.1}h^{-1}\hbox{M}_\odot$, which gives rise to the model predictions that match the simulation mock results well.

\begin{figure*}
\centering
\includegraphics[width=1.0\columnwidth]{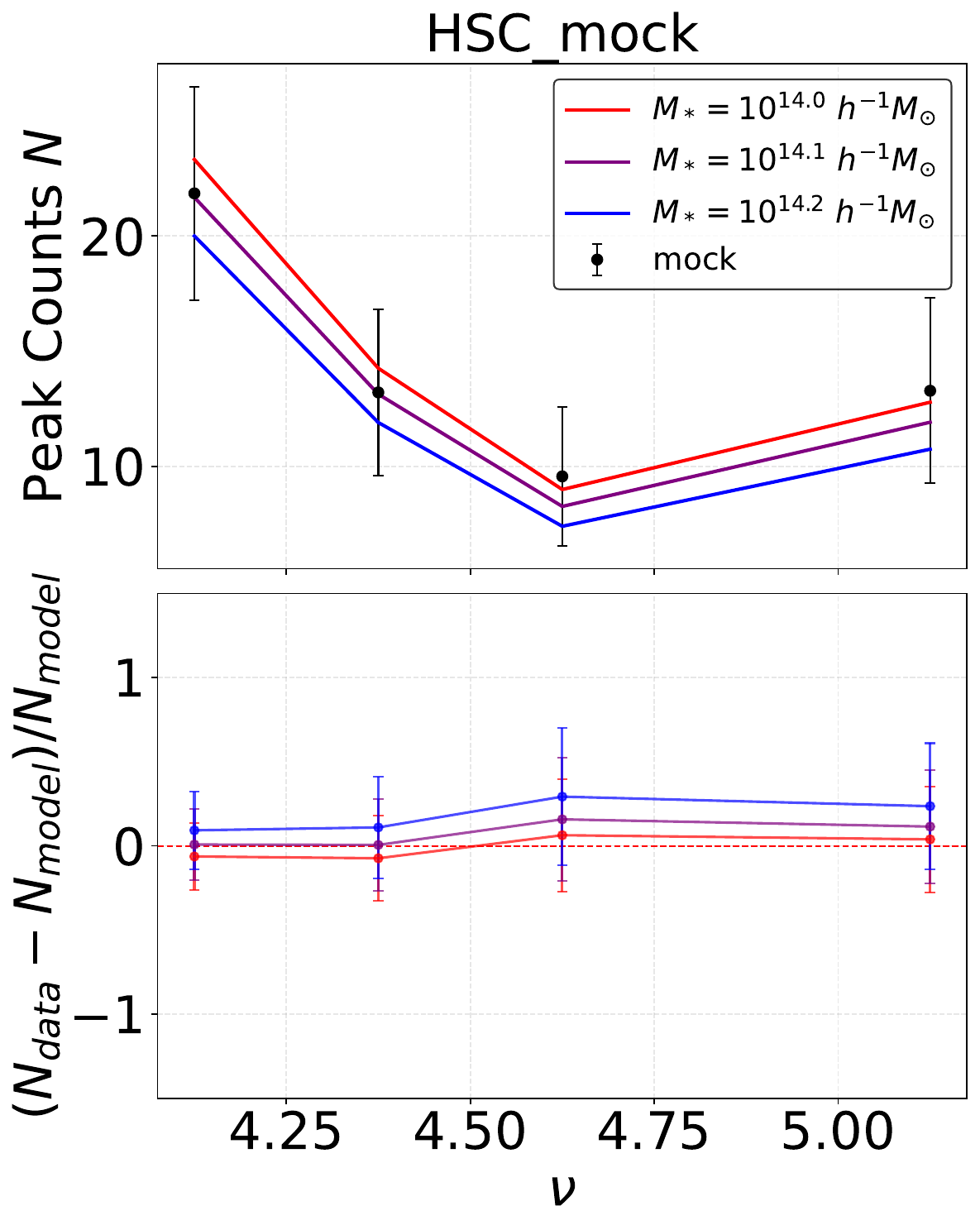}
\includegraphics[width=1.0\columnwidth]{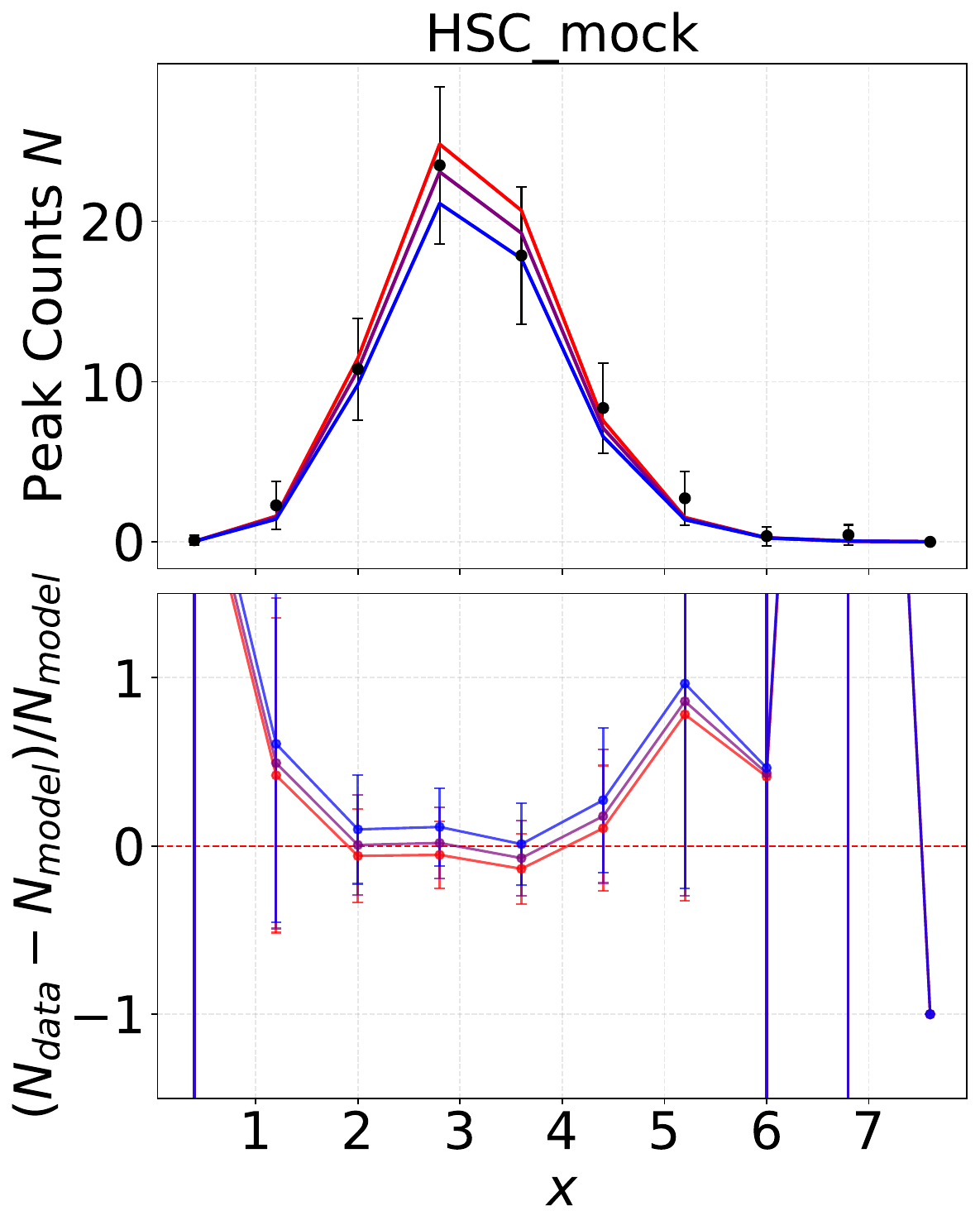}
\includegraphics[width=1.0\columnwidth]{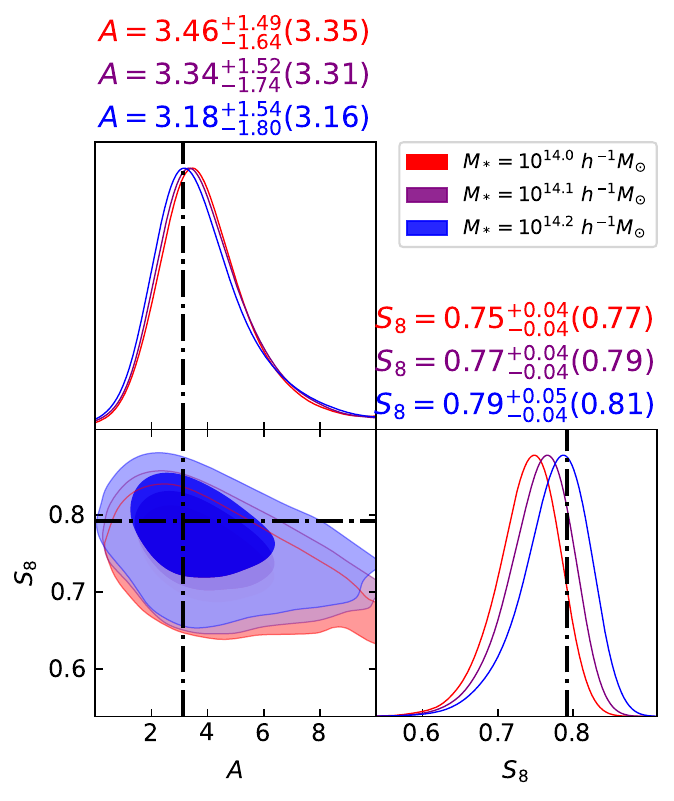}
\includegraphics[width=1.0\columnwidth]{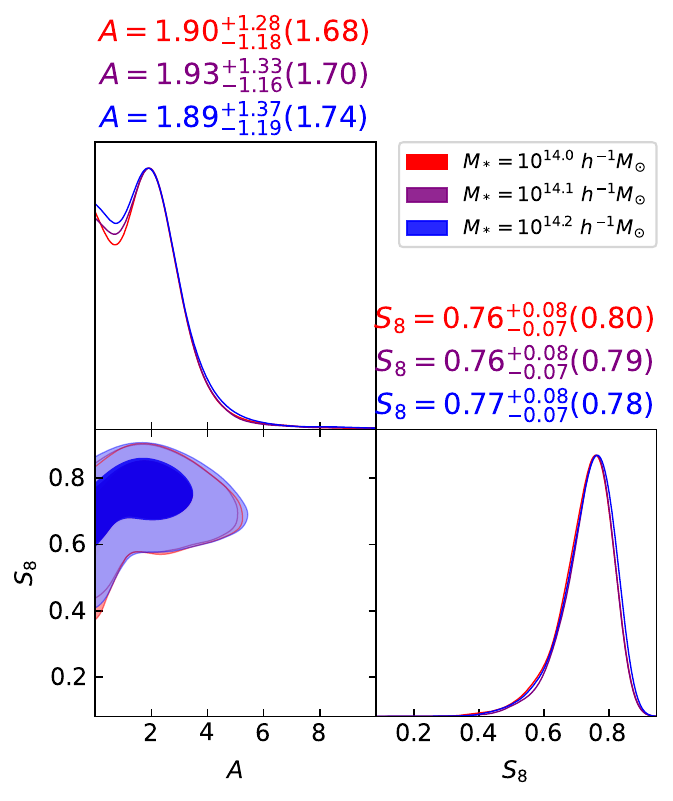}
\caption{\label{fig:M_start} Upper: The average peak counts from our HSC mocks similar as the upper panel of Figure \ref{fig:HSC_mock}.
The different colored lines are the model predictions with different $M_*$.
Lower: The corresponding constraints from the mock data (left) and observational data (right) with the joint peak height and steepness statistics.}
\end{figure*}

Here we test the dependence of our results on the specific choice of $M_*$ and consider $M_*=10^{14.0}h^{-1}\hbox{M}_\odot$, $10^{14.1}h^{-1}\hbox{M}_\odot$ and $10^{14.2}h^{-1}\hbox{M}_\odot$, respectively. The upper panels of Figure \ref{fig:M_start} show the peak height (left) and steepness (right) distributions from the model predictions with different $M_*$ in comparison with the simulation mock data. It is seen that the differences from different $M_*$ calculations are all within $1\sigma$ statistical uncertainties.

The bottom left panel of Figure \ref{fig:M_start} shows the corresponding constraints from mock data and they are consistent within $1\sigma$ range. 
In the bottom right panel, we further present the results from the observed peaks. Again, the different choice of $M_*$ does not affect the constraints significantly. 

\section{Likelihood and covariance\label{app:likelihood}}
In our analyses here, we count the number of peaks and bin them based on the signal-to-noise ratio of height and steepness. Such a counting process is similar to the cosmological studies using cluster counts. Different likelihood forms have been employed in cluster cosmology studies \citep[e.g.,][]{2016A&A...594A..24P,2022A&A...665A.100L,2022A&A...659A..88L,2023MNRAS.520.6223P,2025OJAp....8E...2C}. In these counting cases, given a local average count $N_{\rm {local}}$, the observed ones are the realizations from a Poisson sampling process, which can be approximated by Gaussian realizations if $N_{\rm{local}}$ is large enough. For $N_{\rm{local}}$ itself, it can be considered as a Gaussian realization from a given global average count $N_{\rm{global}}$ taking into account the cosmic sample variance over the survey area. Thus in general, the likelihood is a Gauss-Poisson Compound (GPC) form \citep[e.g.,][]{2003ApJ...584..702H}. In \cite{2023MNRAS.520.6223P}, in terms of cluster counts, they carefully compare different likelihood forms, namely Poisson, Gaussian and GPC. The results show that the Poisson likelihood generally underestimates the errors on the derived cosmological parameter constraints because of the neglect of the sample variance. On the other hand, the Gaussian likelihood performs similarly well as the GPC likelihood, even in the case with a cluster count less than $5$ in certain bins due to the importance of the sample variance.

\begin{figure*}
\centering
\includegraphics[width=2.0\columnwidth]{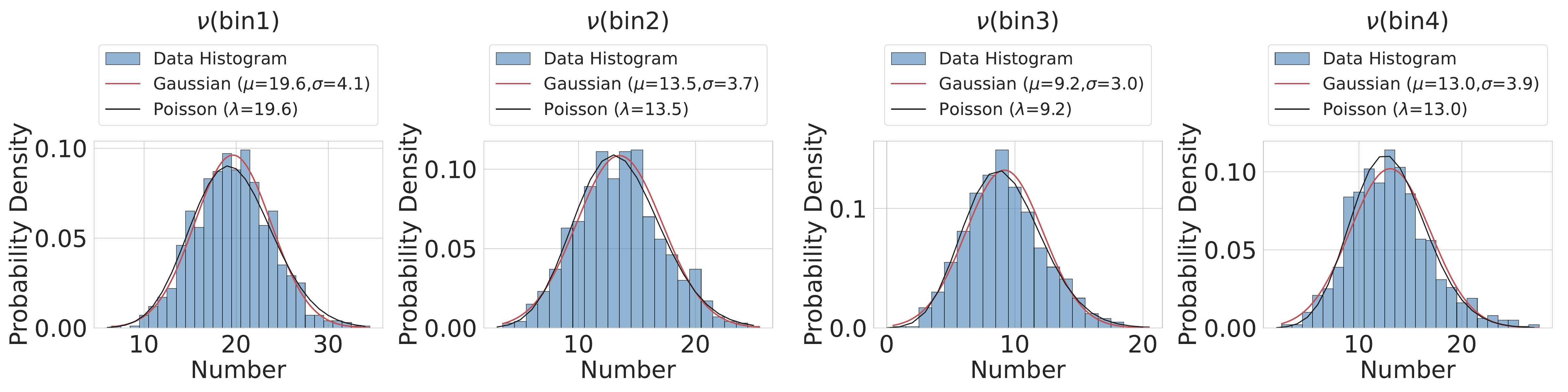}
\includegraphics[width=2.0\columnwidth]{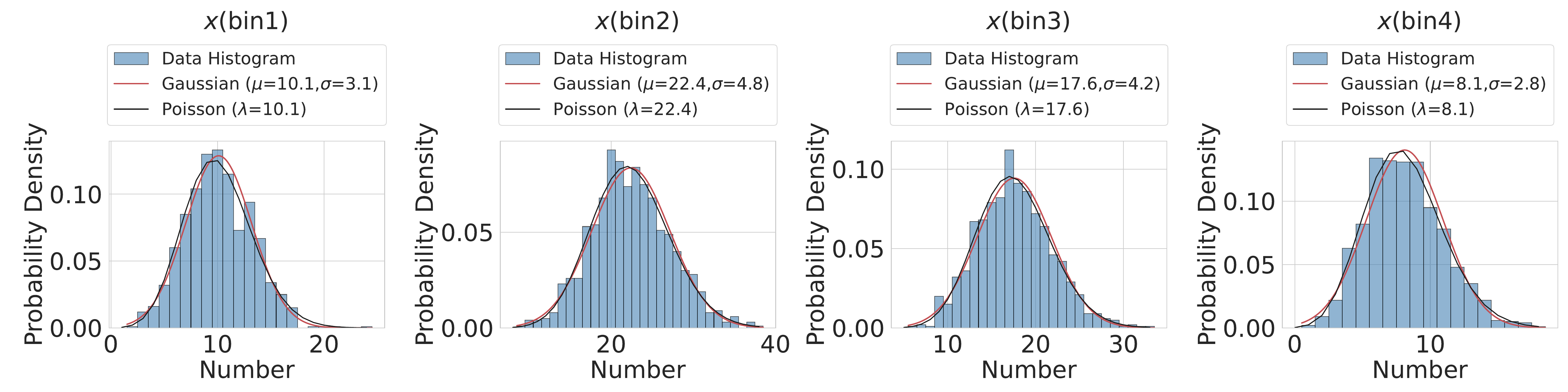}
\caption{\label{fig:bin distributions} Probability density function of the peak counts in each bin. The histogram is derived from the 1000 mock data sets. The black and red lines represent the Poisson and Gaussian distributions, respectively, with the mean and standard deviation estimated from the mock data.} 
\end{figure*}

\begin{figure*}
\centering
\includegraphics[width=1.8\columnwidth]{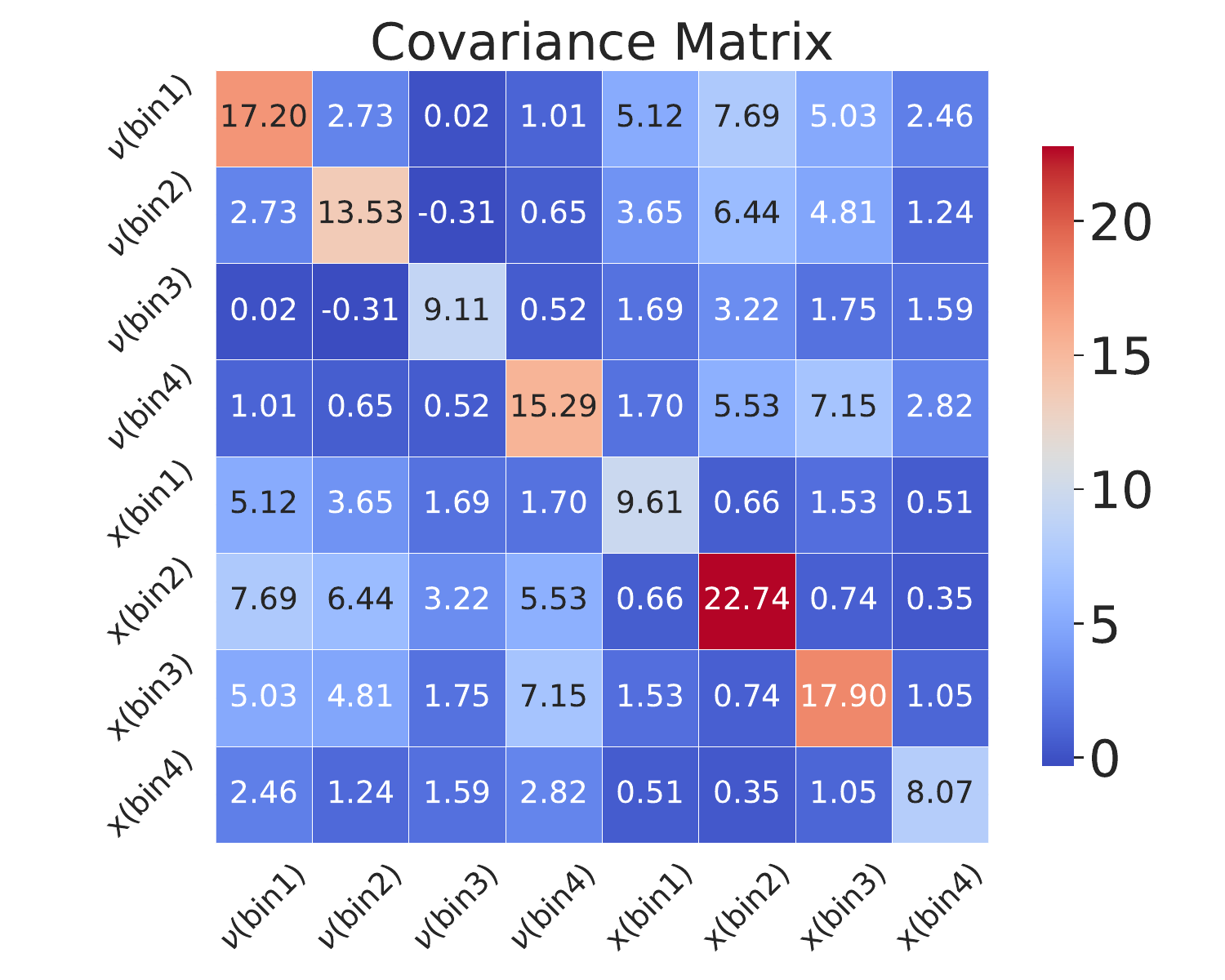}
\caption{\label{fig:covariance} The joint covaraince of peak height and steepness from the 1000 mocks. 
}
\end{figure*}

In our studies, we calculate the distributions of the peak counts in the considered peak height and steepness bins from $1000$ HSC mocks shown in the upper and lower panels of Figure~\ref{fig:bin distributions}, respectively. The red and black lines are the Gaussian and Poisson distributions, respectively. It is seen that the two models describe the histograms similarly well. We also calculate the covariance between the 8 data points from the $1000$ mocks, and the result is shown in Figure~\ref{fig:covariance}. While the diagonal terms are dominant, the off-diagonal ones are not negligible showing the effects of sample variance. We therefore, like many other WL peak analyses \citep[e.g.,][]{2015MNRAS.450.2888L,2016MNRAS.463.3653K,2018MNRAS.474.1116S,2018MNRAS.474..712M,2022MNRAS.511.2075Z,2023MNRAS.519..594L,2024MNRAS.528.4513M, 2024MNRAS.534.3305H} adopt the Gaussian distributed data in our analyses with the logarithmic of a Gaussian likelihood given by
\begin{equation}
\chi_{\bm p}^{2}=\sum_{i j} d {N}_{i}^{\left(d, t\right)}\widehat{\left(C^{(f)}\right)_{i j}^{-1}} d {N}_{j}^{\left(d, t\right)},\label{loglikelihood}
\end{equation}
where $d {N}_{i}^{\left(d, t\right)}={N}_{i}^{\left(d\right)}-{N}_{i}^{\left(t\right)}$ is the difference of the peak number in $i$th bin between data (mock or observational) and
the theoretical prediction given a set of cosmological parameters $\bm p$.
The $\widehat{(C^{(f)})^{-1}}$ is the inverse covariance matrix calculated by an unbiased estimator \citep{2007A&A...464..399H}
\begin{equation}
\widehat{(C^{(f)})^{-1}}=\frac{R-N_{\rm {bin }}-2}{R-1}(C^{(f)})^{-1},
\label{inverse}
\end{equation}
where $(C^{(f)})^{-1}$ is the inverse of the covariance $C^{(f)}$ calculated from the $R=1000$ mocks and $N_{\rm {bin }}$ is the number of bins for peak counts.

It is noted that considering the uncertainties of the covariance matrix estimation, the likelihood of Gaussian distributed data should be a modified version of a multivariate t-distribution that is approaching the Gaussian likelihood when the number of mocks $R$ used to estimate the covariance matrix is large \citep[e.g.][]{2016MNRAS.456L.132S, 2022MNRAS.511.2075Z,2024MNRAS.534.3305H}. With $R=1000$ here, we do not take into account the modified likelihood distribution. In our future studies, we will investigate the impacts of different forms of the likelihood on precision WL peak analyses.

\bibliography{HSCpeaksteepness}{}
\bibliographystyle{aasjournalv7}



\end{document}